%% file: DistanceEmbeddings-arXiv.tex
\documentclass[10pt,reqno,final]{amsart}
\pdfoutput=1
\usepackage{fourier,fullpage} 
\usepackage[sort&compress]{natbib}
\setcitestyle{numbers,square,comma}
\usepackage{amsmath,amsfonts,amssymb,amsthm,graphicx,color,bbold,url}

\title{Representation and Coding of Signal Geometry} 

\author{Petros T. Boufounos}
\address{Mitsubishi Electric Research Laboratories\\201 Broadway\\Cambridge
  MA 02139,\\USA}
\email{petrosb@merl.com}
\author{Shantanu Rane}
\thanks{S. Rane performed this work while he was a researcher at MERL}
\address{Palo Alto Research Center\\3333 Coyote Hill Road\\Palo Alto, CA
  94304\\USA}
\email{srane@parc.com}
\author{Hassan Mansour}
\address{Mitsubishi Electric Research Laboratories\\201 Broadway\\Cambridge,
  MA 02139\\USA}
\email{mansour@merl.com}

\input{definitions}

\theoremstyle{plain}
\newtheorem{theorem}{Theorem}[section]

\newtheorem{proposition}[theorem]{Proposition}
\newtheorem{corollary}{Corollary}[section]
\theoremstyle{definition}
\newtheorem{definition}[theorem]{Definition}

\theoremstyle{remark}

\begin{document}


\begin{abstract}
\input{abstract}
\end{abstract}

\maketitle

\input{introduction}

\input{background}

\input{probconstruct}

\input{designandanalysis}

\input{examples}

\input{simulations}

\input{discussion}
\appendix
\input{app_proofcontinuous}
\input{app_proofdiscontinuous}
\input{app_proofdesign}
\input{app_proofsubadditive}
\input{app_proofbinarycollisionbounds}

\bibliographystyle{siam}
\bibliography{references}

\end{document}

%% file: definitions.tex
\newcommand{\mA}{\ensuremath{\mathbf A}}

\newcommand{\vy}{\ensuremath{\mathbf y}}
\newcommand{\vz}{\ensuremath{\mathbf z}}
\newcommand{\vx}{\ensuremath{\mathbf x}}

\newcommand{\va}{\ensuremath{\mathbf a}}
\newcommand{\vq}{\ensuremath{\mathbf q}}

\newcommand{\vu}{\ensuremath{\mathbf u}}
\newcommand{\vw}{\ensuremath{\mathbf w}}
\newcommand{\ball}[2]{\ensuremath{\mathcal{B}_{{#1}}({{#2}})}}
\newcommand{\sW}{\ensuremath{\mathcal{W}}}
\newcommand{\sS}{\ensuremath{\mathcal{S}}}

\newcommand{\Reals}{\ensuremath{\mathbb{R}}}

\newcommand{\sign}{\ensuremath{\mathrm{sign}}}

%% file: keywords.tex
Randomized Embeddings, Dimensionality Reduction, Distance representations, Coding for
  inference, Kernel methods, Quantization

%% file: abstract.tex
Approaches to signal representation and coding theory have traditionally focused on how
to best represent signals using parsimonious representations that incur the
lowest possible distortion. Classical examples include linear and
non-linear approximations, sparse representations, and rate-distortion
theory. Very often, however, the goal of processing is to extract
specific information from the signal, and the distortion should be
measured on the extracted information. The corresponding
representation should, therefore, represent that information as
parsimoniously as possible, without necessarily accurately
representing the signal itself.

In this paper, we examine the problem of encoding signals such that
sufficient information is preserved about their pairwise distances and
their inner products. For that goal, we consider randomized embeddings
as an encoding mechanism and provide a framework to analyze their
performance. We also demonstrate that it is possible to design the
embedding such that it represents different ranges of distances with
different precision. These embeddings also allow the computation of
kernel inner products with control on their inner product-preserving
properties. Our results provide a broad framework to design and
analyze embeddins, and generalize existing results in this area, such
as random Fourier kernels and universal embeddings.

%% file: introduction.tex
\section{Introduction}
\label{sec:intro}
Signal representation theory and practice has primarily focused on how
to best represent or encode a signal while incurring the smallest
possible distortion. For example, image or video representations
typically aim to minimize the distortion in the signal so that the
visual quality of the signal is maintained when displayed to a
user. Quite often, however, the user of a signal is not a human
observer, but an algorithm extracting some information about the signal. 
In this case, the goal is different: the representation should not destroy the
information that the algorithm requires, even if the signal itself
cannot be completely recovered.

In this paper we examine signal representations that preserve aspects
of the signal's geometry
but not necessarily the signal itself. Our approach exploits the
geometry-preserving properties of randomized
embeddings. Specifically, we develop a framework that generalizes
well-known embeddings in a manner than enables the design and control
of the distance distortion and the resulting inner product kernel. The
results in this paper extend and generalize recently developed theory
for efficient universal quantization and universal quantized
embeddings~\cite{B_TIT_12,BR_WIFS11,BR_DCC13,BM_SAMPTA2015} and random
Fourier kernels~\cite{Rahimi07,raginsky2009locality}. We demonstrate
and analyze such representations using a very general approach, that
can encompass continuous and quantized embeddings.

As we first reported in~\cite{BR_DCC13,BM_SAMPTA2015}, representations
based on universal embeddings---which are special cases of our
development---can be used as a geometry-preserving coding mechanism in
image retrieval and classification applications. We demonstrated that
we are able to improve compression performance up to $25\%$ over
previous embedding-based approaches~\cite{CP,RP}, including our own
earlier work~\cite{LRB_MMSP12}. The main advantage of our approach is
the ability to control the range of distances best preserved by the
embeddings, so that we do not represent distance ranges that are not
important to the application at hand. In most inference applications
it is only necessary to represent distances up to a certain radius, as
required by the algorithm, and not any farther. Thus, bits are not
wasted in coding distances larger than necessary.

\input{in_motivation}

\input{in_contributions}

\input{in_relatedwork}

\input{in_notation}

\input{in_outline}

%% file: in_motivation.tex
\subsection{Motivation}
\label{sec:motivation}
Our work is partly motivated by cloud-based image retrieval and
classification applications, such as augmented reality. As we discuss
in~\cite{LRB_MMSP12,BR_DCC13,BM_SAMPTA2015}, augmented reality and
other image retrieval and classification applications can benefit
significantly by efficient coding of the geometry of the signal space
and of the geometric relationships between signals.

In typical cloud-based image retrieval applications, a client
transmits to a cloud server a query image acquired by the user, or
features extracted from that image, requesting more information on the
objects in the image. The server extracts features, if necessary, and
uses those to execute the query. The query typically drives a
machine learning algorithm, such as nearest neighbor search or support
vector machines (SVMs), searching the database for metadata associated
with those features. The query typically returns information such as
the object class, the object identity, or associated information about
the object. This search should be computationally efficient
at the client and the database server, 
and the transmission should be bandwidth-efficient.

While our example application uses nearest-neighbor search and SVMs for
classification and inference, our goal is to develop a
framework that is agnostic to the underlying mechanism, as long as
this mechanism relies on signal geometry for its functioning. Thus, most
classical machine learning techniques, such as regressions, mixture
models, spectral clustering and deep learning, can immediately exploit
our representations. All that is necessary in order to appropriately
design the embedding is an understanding of which signal distances
should be preserved for the particular application.

Our hope is that this analysis framework will become a 
useful addition to a system designer's tool-belt, providing methods that can 
be layered with other machine learning primitives. 
In this paper, we concentrate on one particular method for designing 
geometry-preserving embeddings and develop 
a general method to analyze them. However, some of the theoretical
tools developed here can be used in more general settings and
applications.

%% file: in_contributions.tex
\subsection{Contributions}
Our paper contributes several results toward establishing embeddings
as general representations of signal geometry:
\begin{itemize}
\item We introduce a generalized definition and characterization of
  geometry-preserving embeddings which allows for selective
  distortions in the signal geometry. These distortions are captured
  using a distance map that describes how distances in the original
  signal space are distorted in the embedding space. This definition
  covers a large number of existing embeddings, as well as more
  general designs, and enables analysis of the embedding
  characteristics given the distance map.
\item We develop a very general framework to extend embeddings to
  infinite sets, such as sparse signals or manifolds, even if the
  mapping function is discontinuous. Our approach is fundamentally
  very similar to established approaches using set coverings. However,
  these methods fail if the embedding function is not continuous,
  e.g., due to quantization. The tools we introduce extend the notion
  of Lipschitz continuity to a large variety of discontinuous
  functions in a way that enables proofs using covering arguments.
\item We demonstrate a method to design randomized embeddings such
  that they achieve the desired distortions in the geometry of the
  space. The design we describe generalizes existing embedding
  constructions, such as the random Fourier features~\cite{Rahimi07},
  and universal quantized
  embeddings~\cite{B_TIT_12,BR_WIFS11,BR_DCC13,BM_SAMPTA2015}.
\item We present an analysis of the embedding ambiguity in the context
  of the distance map. We characterize this ambiguity from a new
  perspective: we assume the embedding is used as a representation of
  the original signal set. Current embedding guarantees describe the
  ambiguity in the embedding space, as opposed to the signal
  set. While the two are equivalent in many well-known cases, they
  differ quite often, especially if the embedding distorts the signal
  geometry.
\item We establish a connection between distance embeddings and kernel
  methods, demonstrating that the distortion of the distance map
  performed by the embedding is equivalent to the distortion performed
  by a kernel inner product.
\item We provide an analysis of multibit universal embeddings and a
  generalization of binary universal embeddings to infinite sets, such
  as sparse signals or manifolds. These generalizations establish new results 
  in this area and serve as 
  examples of how our developments can be used in practice.
\end{itemize}

%% file: in_relatedwork.tex
\subsection{Related Work}
The best known embeddings are due to Johnson and Lindenstrauss
(J-L)~\cite{JL}, which preserve $\ell_2$ distances of point clouds. A
significant body of work has been devoted to developing such
embeddings using a variety of randomizations and for a variety of
applications~\cite{dasgupta03rsa,achlioptas03css}. Their importance
was re-established recently thanks to the emergence of compressive
sensing
(CS)~\cite{CanRomTao::2006::Stable-signal,Can::2006::Compressive-sampling,Don::2006::Compressed-sensing}. The
Restricted Isometry Property (RIP), which plays a central role in CS
theory, is essentially a restatement of the JL property, but applied
to unions of signal subspaces instead of point
clouds~\cite{BarDavDeV::2008::A-Simple-Proof,CandesRIP}. Consequently,
several connections between the two have been established. In addition
to the RIP, extensions of the J-L lemma to other infinite sets, such a
manifolds~\cite{baraniuk2009random} and unions of
subspaces~\cite{blumensath2009sampling,EldMis::2009::Robust-recovery,baraniuk2010model},
have also been established.

Significant literature has also studied variations of J-L
embeddings. For example, in a number of acquisition systems and coding
applications, it is necessary to quantize the
representations. Quantized J-L embeddings have been well
studied~\cite{LRB_MMSP12,Jacques15buffon,jacques2015small}, especially
down to 1-bit per representation
coefficient~\cite{JLBB_1bit_2011,plan2013one,plan2013robust,plan2014dimension}. Furthermore,
while J-L embeddings and the RIP preserve $\ell_2$ distances, there is
a large body of work in preserving other similarity measurements, such
as $\ell_p$ distances for various
$p$'s~\cite{indyk2006stable,jacques2011dequantizing,jacques2013stabilizing,oymak2015near},
edit
distance~\cite{levenshtein66doklady,andoni03siam,bar04focs,ostrovsky07acm}
and angle, i.e., correlation, between
signals~\cite{B_SampTA13,B_SPARS13,B_SPIE2013_APQPE}.

A common thread in the aforementioned body of work is that distances
or other similarity measures are preserved indiscriminately. This is
in sharp contrast to our work, which allows the design of embeddings
that represent some distances better than others, with control on that
design. For example, in our motivating applications in the area of
image retrieval, we design embeddings that only encode a short range
of distances, as necessary for nearest-neighbor computation and
classification. A very narrow notion of locality was discussed in very
recent work, fit for the development in that
paper~\cite{oymak2015near}. That definition, however, does not capture
the richer set of locality properties presented in our line of work.

Recent work has also provided classification guarantees for J-L
embeddings~\cite{bandeira2014compressive} on very particular signal
models. In particular, it is shown that separated convex ellipsoids
remain separated when randomly projected to a space with sufficient
dimensions. Our work significantly enhances the available design space
compared to J-L embeddings. It should, thus, be possible to establish
similar results. However, it is not clear that the techniques
in~\cite{bandeira2014compressive} can be used with our designs. Thus,
establishing results of similar type remains an interesting problem.

Many of our proof techniques rely on
well-established concentration of measure arguments and methods common
in the embedding literature, e.g.,
see~\cite{dasgupta03rsa,achlioptas03css,BarDavDeV::2008::A-Simple-Proof,JLBB_1bit_2011}. However,
we provide a new approach to handle quantization or other
discontinuous distortions, which can significantly expand the
applicability of established approaches. Our main novelty in computing
the embedding is the introduction of a \emph{non-linear, periodic distortion}
that enables notable control over the behavior of the embedding. We also
develop a framework to analyze the performance of the embedding in
preserving distances which, in contrast to the existing literature,
takes into consideration the distortion as manifested in the original
signal distance, as opposed to the embedded distance.

Our work is also related to locality-sensitive hashing (LSH) methods,
which significantly reduce the computational complexity of
near-neighbor computation~\cite{indyk98stoc,Datar04_LSH,LSH}. The LSH
literature shares a lot of the tools with the embeddings literature,
such as randomized projections, dithering and quantization, but the
goal is different: given a query point, LSH will return its near
neighbors very efficiently, with $O(1)$ computation. This
efficiency comes at a cost: no attempt is made to represent the
distances of neighbors. When used to compare signals it only provides
a binary decision, whether the distance of the signals is smaller than
a threshold or not. This makes it unsuitable for applications that
require more accurate distance information.

Our design does not explicitly take retrieval complexity into account;
we expect the underlying retrieval machinery to consider complexity
issues. Nevertheless, our methods provide dimensionality and bit-rate
reduction, which are tightly coupled to complexity. Furthermore, some
of our embedding techniques could be used in the context of an
LSH-based scheme; some of the LSH techniques
in~\cite{indyk98stoc,Datar04_LSH,LSH} are reminiscent of our
approach. It should also be possible to design mechanisms that reduce
complexity which explicitly exploit our methods, for example extending
the hierarchical approach
in~\cite{B_SampTA11}. However, such designs, although
quite interesting, are beyond the scope of this paper.

Our work is of similar flavor to~\cite{Rahimi07,raginsky2009locality},
which use randomized embeddings to efficiently approximate specific
kernel computations. The results we present generalize these
approaches, by allowing control over the distance map in the kernel and
the ambiguity of the distance preservation. We further provide a
general approach to understand the approximation properties of the
embedding and its behavior under quantization. 

There is also a large body of work focused on learning embeddings from
available data, e.g.,
see~\cite{SpectralH,LDAHash,HSYB_TSP15_NuMax,sadeghian2013energy}. Such
approaches exploit a computationally expensive training stage to
improve embedding performance with respect to its distance-preserving
properties. Still, the embedding guarantees are only applicable to
data similar to the training data; the embedding might not perform
well on different sets. Instead, our approach relies on a
randomization independent of the data. Our designs are universal in
the sense that they work on any data set with overwhelming
probability, as long as the embedding parameters are drawn
independently of the dataset. Of course, using data for training is a
promising avenue and a potentially useful extension of our
work. However, we do not attempt this in this paper.

%% file: in_notation.tex
\subsection{Notation}
In the remainder of the paper we use regular typeface, e.g., $x$ and
$y$, to denote scalar quantities. Lowercase boldface such as
\vx\ denotes vectors and uppercase boldface such as \mA\ denotes
matrices. Functions are denoted using regular lowercase typefaces,
e.g., $g(\cdot)$. Unless explicitly noted, all functions are scalar
functions of one variable. In abuse of notation, a vector input to
such functions, e.g., $g(\vx)$ means that the function is applied
element-wise to all the elements of \vx. Sets and vector
spaces are denoted using calligraphic fonts, e.g., \sW, \sS. 

The Fourier transform of a function $h(x)$ is defined as
$H(\xi)=\int_{-\infty}^{+\infty} h(x) e^{-2\pi \mathbb{i}x\xi}$, where
$\mathbb{i}=\sqrt{-1}$ is the imaginary unit. Similarly, the
characteristic function of a probability density function $f_x(x)$ is
defined as
$\phi_x(t)=E\left[e^{\mathbb{i}tx}\right]=\int_{-\infty}^{+\infty}
f_x(x) e^{\mathbb{i}x\xi}$. Thus, the Fourier transform of the density
is related to its characteristic function:
$F_x(\xi)=\phi_x(-2\pi\xi)=\phi_x^*(2\pi\xi)$, where $(\cdot)^*$
denotes complex conjugation. Conditional distributions and
characteristic functions are denoted using $\cdot|\cdot$ in their
argument, e.g., $f_x(x|y)$ and $\phi_x(x|y)$.

%% file: in_outline.tex
\subsection{Outline}
The next section contains a brief background on embeddings and
universal scalar quantization, establishing notation and
definitions. It also reviews Lipschitz continuity and introduces a
generalization that will prove very useful in our subsequent
development, especially for quantized
embeddings. Section~\ref{sec:prob_construction} provides an overview
of how embedding results are typically established on point clouds
using concentration of measure arguments, and introduces our framework
to generalize such embeddings---both quantized and unquantized
ones---to continuous sets. Section~\ref{sec:design_analysis}
demonstrates that embeddings and embedding maps can be designed to
preserve different ranges of distances with different accuracy by
establishing such a design, as well as the tools to analyze its
properties.

Examples of embeddings established using our tools are provided in
Sec.~\ref{sec:examples}. These include quantized Johnson-Lindenstrauss
embeddings, and binary and multibit universal embeddings. These
examples demonstrate how our tools can be applied. They also establish
some new useful results for universal embeddings. In addition,
Sec.~\ref{sec:simul-appl-exampl} provides simulations and application
examples that validate our theory and demonstrate how it can be used
in practice. Finally, Sec.~\ref{sec:discussion} discusses our results
and concludes. For most of our results, in order to improve the flow
and readability of our paper, we have relegated the proofs to
appendices.

%% file: background.tex
\section{Definitions and Background}
\label{sec:bg}
\input{bg_randembeddings}
\input{bg_universalembeddings}
\input{bg_continuity}

%% file: bg_randembeddings.tex
\begin{figure*}[t]
\begin{minipage}[b]{.45\linewidth}
  \begin{center}
    (a)~~
    \includegraphics[width=.9\linewidth]{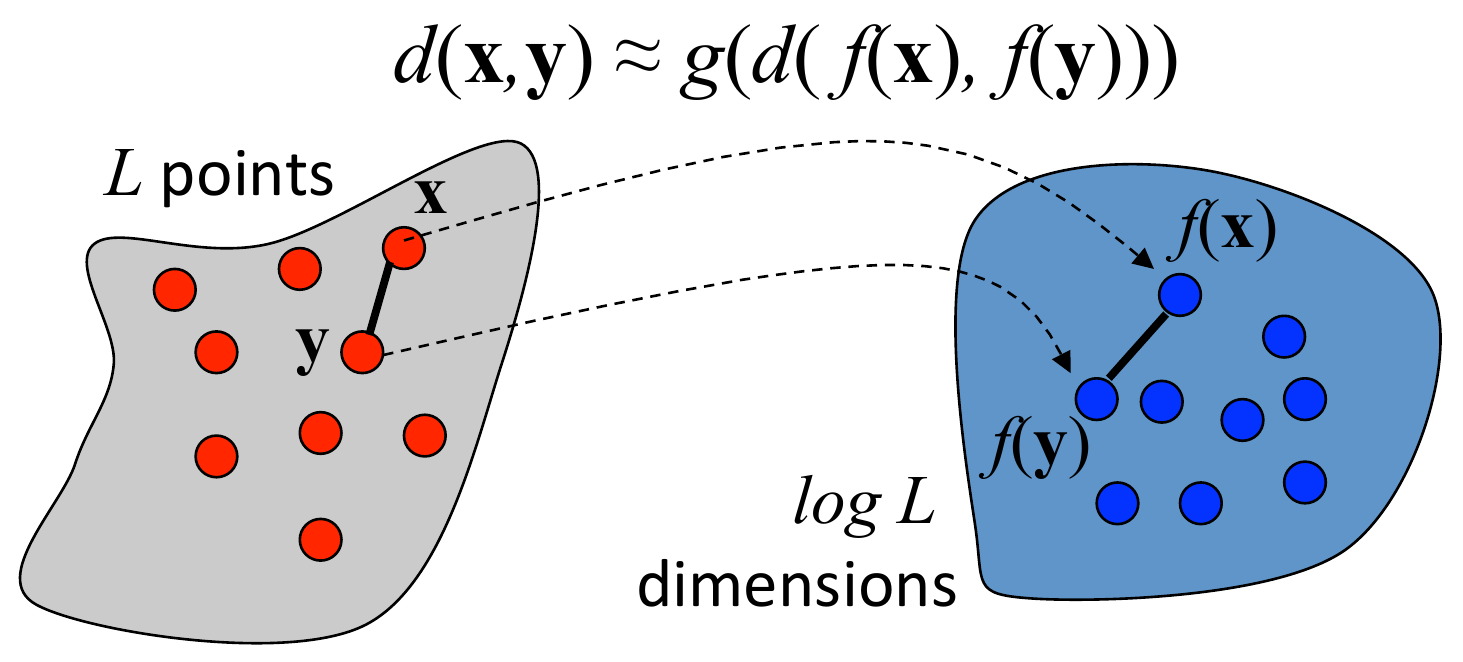}
  \end{center}
\end{minipage}
\hfill
\begin{minipage}[b]{.45\linewidth}
  \begin{center}
    (b)~~
    \includegraphics[width=.9\linewidth]{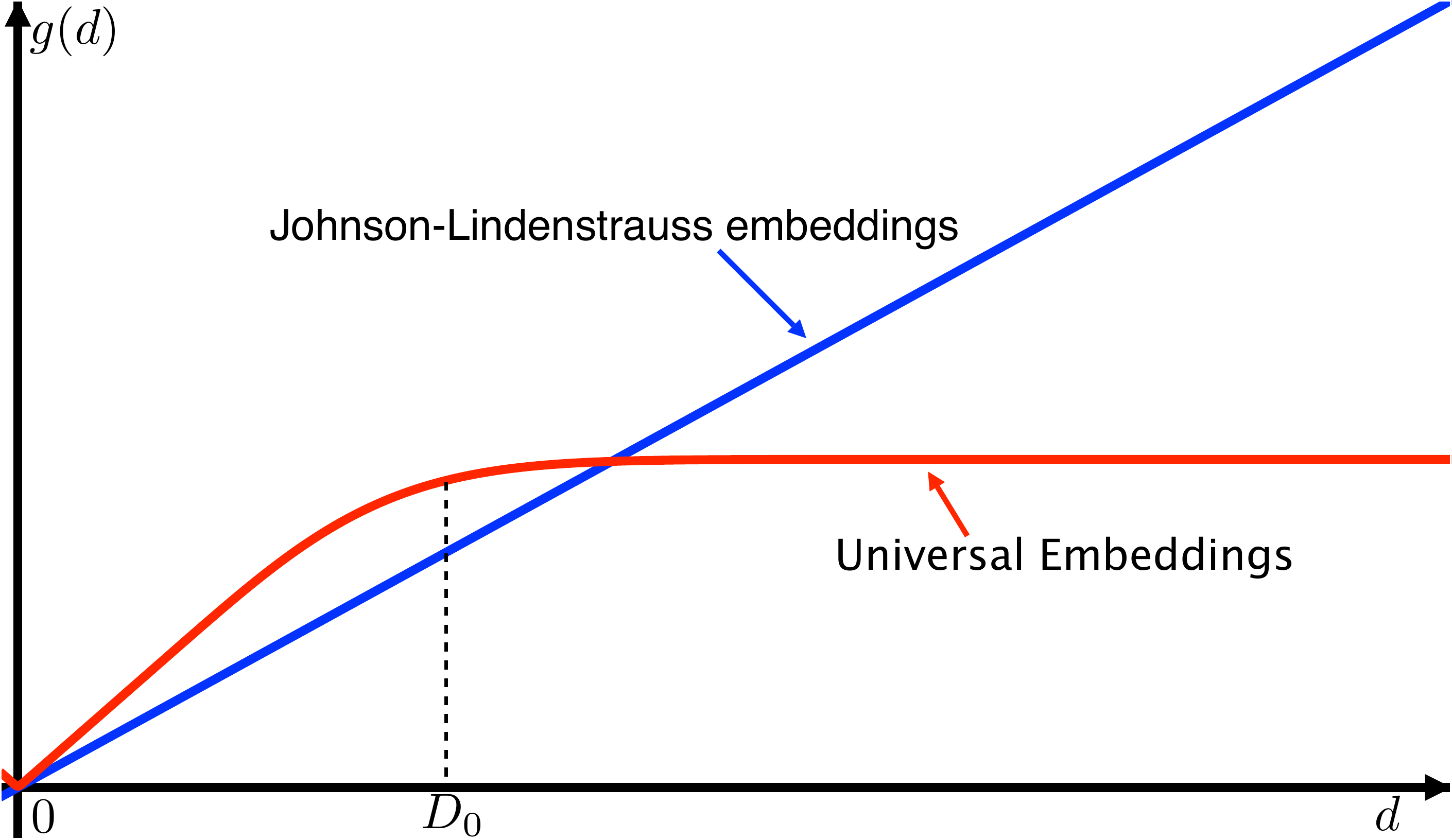}
  \end{center}
\end{minipage}
\caption{(a) Distance-preserving embeddings approximately preserve a
  function $g(\cdot)$ of the distance, allowing distances to be
  computed in a space that (typically) has fewer dimensions and
  produce signals that often require lower transmission rate. (b) For
  most embeddings, such as JL Embeddings, this function is linear, as
  shown in blue. For the universal quantized embeddings discussed in
  this paper, the function is approximately linear initially and
  quickly flattens out after a certain distance $D_0$, as shown in red.}
\label{fig:JL}
\end{figure*}

\subsection{Randomized Linear Embeddings}
An embedding is a transformation of a set of signals in a
high-dimensional space to a (typically) lower-dimensional one such
that some aspects of the geometry of the set are preserved, as
depicted in Fig.~\ref{fig:JL}(a). Since the set geometry is preserved,
distance computations can be performed directly on the
low-dimensional---and often low bitrate---embeddings, rather than the
underlying signals. For the purposes of this paper, we define an
embedding as follows.
\begin{definition}
\label{def:general_embedding}
A function $f:\sS\rightarrow\sW$ is a $(g,\delta,\epsilon)$ embedding
of \sS\ into \sW\ if, for all $\vx,\vy\in\sS$, it satisfies
\begin{align}
  (1-\delta)g\left(d_{\sS}(\vx,\vy)\right)-\epsilon
  \le d_{\sW}\left(f(\vx),f(\vy)\right)\le
  (1+\delta)g\left(d_{\sS}(\vx,\vy)\right)+\epsilon.
  \label{eq:general_embedding}
\end{align}
\end{definition}

In this definition $g:\Reals\rightarrow\Reals$ is an invertible
function mapping distances in $\sS$ to distances in $\sW$ and $\delta$
and $\epsilon$ quantify, respectively, the multiplicative and the
additive ambiguity of the map\footnote{Often it makes more sense to
  separate the upper and lower multiplicative bounds $(1\pm\delta)$ to
  different constants $A,B$. This does not affect the subsequent
  development but encumbers the notation, so we avoid it in this
  paper.}. We will often refer to $g(\cdot)$ as the distance map and
to $f(\cdot)$ as the embedding map. In most known embeddings, such as
the ones discussed in this section, the distance map is the identity
$g(d)=d$ or a simple scaling.

The best known embeddings are the Johnson-Lindenstrauss
embeddings~\cite{JL}. These are functions $f:\sS\rightarrow \Reals^M$
from a finite set of signals $\sS\subset\Reals^N$ to a $M$-dimensional
vector space such that, given two signals $\vx$ and $\vy$ in \sS,
their images satisfy:
\begin{align}
  (1-\delta)\|\vx-\vy\|_2^2\le\|f(\vx)-f(\vy)\|_2^2\le(1+\delta)\|\vx-\vy\|_2^2.
  \label{eq:JLembedding}
\end{align}
In other words, these embeddings preserve Euclidean, i.e., $\ell_2$,
distances of point clouds within a small factor, measured by $\delta$,
and using the identity as a distance map.

Johnson and Lindenstrauss demonstrated that a distance-preserving embedding,
as described above, exists in a space of dimension
$M=O(\frac{1}{\delta^2}\log L)$, where $L$ is the number of signals in
\sS\ (its cardinality) and $\delta$ the desired tolerance in the
embedding. Remarkably, $M$ is independent of $N$, the dimensionality
of the signal set \sS. Subsequent work showed that it is
straightforward to compute such embeddings using a linear mapping. In
particular, the function $f(\vx) = \mathbf{A}\vx$, where $\mathbf{A}$
is a $M\times N$ matrix whose entries are drawn randomly from specific
distributions, satisfies~\eqref{eq:JLembedding} for all
$\vx,\vy\in\sS$ with probability $1-c_1e^{\log L-c_2\delta^2M}$, for
some universal constants $c_1,c_2$, where the probability is with
respect to the measure of \mA. Commonly used distributions for the
entries of \mA~are i.i.d. Gaussian, i.i.d. Rademacher, or
i.i.d. uniform~\cite{dasgupta03rsa,achlioptas03css}.

More recently, in the context of compressive sensing, such linear
embeddings have been shown to embed infinite sets of signals. For
example, the restricted isometry property (RIP) is an embedding of
$K$-sparse signals and has been shown to be achievable with $M=O(K\log
N/K)$~\cite{BarDavDeV::2008::A-Simple-Proof,CandesRIP}. Similar
properties have been shown for other signal set models, such as more
general unions of
subspaces~\cite{blumensath2009sampling,EldMis::2009::Robust-recovery,baraniuk2010model}
and manifolds~\cite{baraniuk2009random}. Typically, these
generalizations are established by first proving that the embedding
holds in a sufficiently dense point cloud on the signal set and
exploiting linearity and smoothness to extend it to all the points of
the set.

Such embeddings result in a significant dimensionality
reduction. However, dimensionality reduction does not immediately
produce rate reduction; the embeddings must be quantized for
transmission and, if the quantization is not well designed,
performance suffers \cite{LRB_MMSP12}. In particular, when combined
with scalar quantization, the embeddings satisfy
\begin{align}
  (1-\delta)\|\vx-\vy\|_2-\epsilon \le\|f(\vx)-f(\vy)\|_2\le (1+\delta)\|\vx-\vy\|_2+\epsilon,
\end{align}
where $\epsilon\propto 2^{-B}$ is the quantizer step size, decreasing
exponentially with the number of bits used per dimension, $B$. On the
other hand, $\delta$ is a function of $M$, the projection's
dimensionality, and scales approximately as $1/\sqrt{M}$, as is the
case for the J-L embedding. Recent work
has refined these bounds, demonstrating that $\epsilon$ and $\delta$
decrease together as the number of measurements decreases when
considering an $\ell_2$ embedding into
$\ell_1$~\cite{Jacques15buffon,jacques2015small}. In the extreme case
of 1-bit scalar quantization the quantizer only keeps the sign of each
measurement. Thus, a binary embedding does not preserve signal
amplitudes, and therefore, $\ell_2$ distances. Still, it does preserve
angles, or equivalently, correlation
coefficients~\cite{JLBB_1bit_2011,plan2014dimension,plan2013one,plan2013robust}.

When designing a quantized embedding, the total rate is determined by
the dimensionality of the projection and the number of bits used per
dimension: $R=MB$. For a fixed bit budget $R$, as the dimensionality $M$
increases, the accuracy of the embedding before quantization, as
reflected in $\delta$, is increased. But to keep the rate fixed, the
number of bits per dimension should also decrease, which decreases the
accuracy due to quantization, reflected in $\epsilon$. This non-trivial
trade-off is explored in detail in~\cite{LRB_MMSP12}; at a constant
rate a multibit quantizer outperforms the 1-bit quantizers examined in
earlier literature~\cite{CP,RP}.

%% file: bg_universalembeddings.tex
\subsection{Universal Quantization and Embeddings}
\label{sec:univ-quant-embedd}
Universal scalar quantization, first introduced in~\cite{B_TIT_12},
fundamentally revisits scalar quantization and redesigns the quantizer
to have non-contiguous quantization regions. This approach also relies
on a Johnson-Lindenstrauss style projection, followed by scaling,
dithering and scalar quantization:
\begin{align}
f(\vx)=Q(\Delta^{-1}(\mA\vx+\vw)), 
\label{eq:universal_quantization}
\end{align}
where \mA\ is a $M\times N$ random matrix with
$\mathcal{N}(0,\sigma^2)$-distributed, i.i.d. elements,
$\Delta^{-1}$---abusing notation---an element-wise scaling factor,
\vw\ a length-$M$ dither vector with i.i.d. elements, uniformly
distributed in $[0,2^B\Delta]$, and $Q(\cdot)$ a $B$-bit scalar
quantizer operating element-wise on its input.

\begin{figure*}[t]
\begin{minipage}[b]{.3\linewidth}
  \noindent\small{(a)}~~
  \begin{minipage}[b]{.9\linewidth}
    \includegraphics[width=\linewidth]{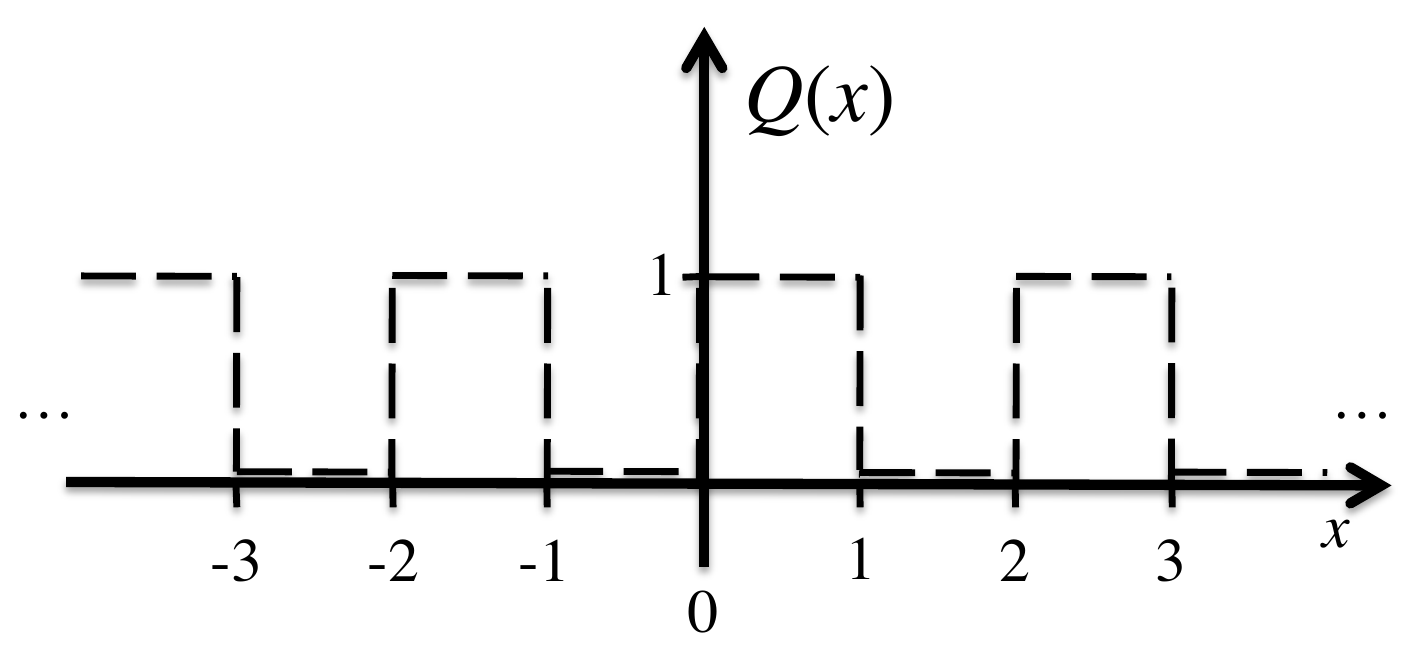}
    \includegraphics[width=\linewidth]{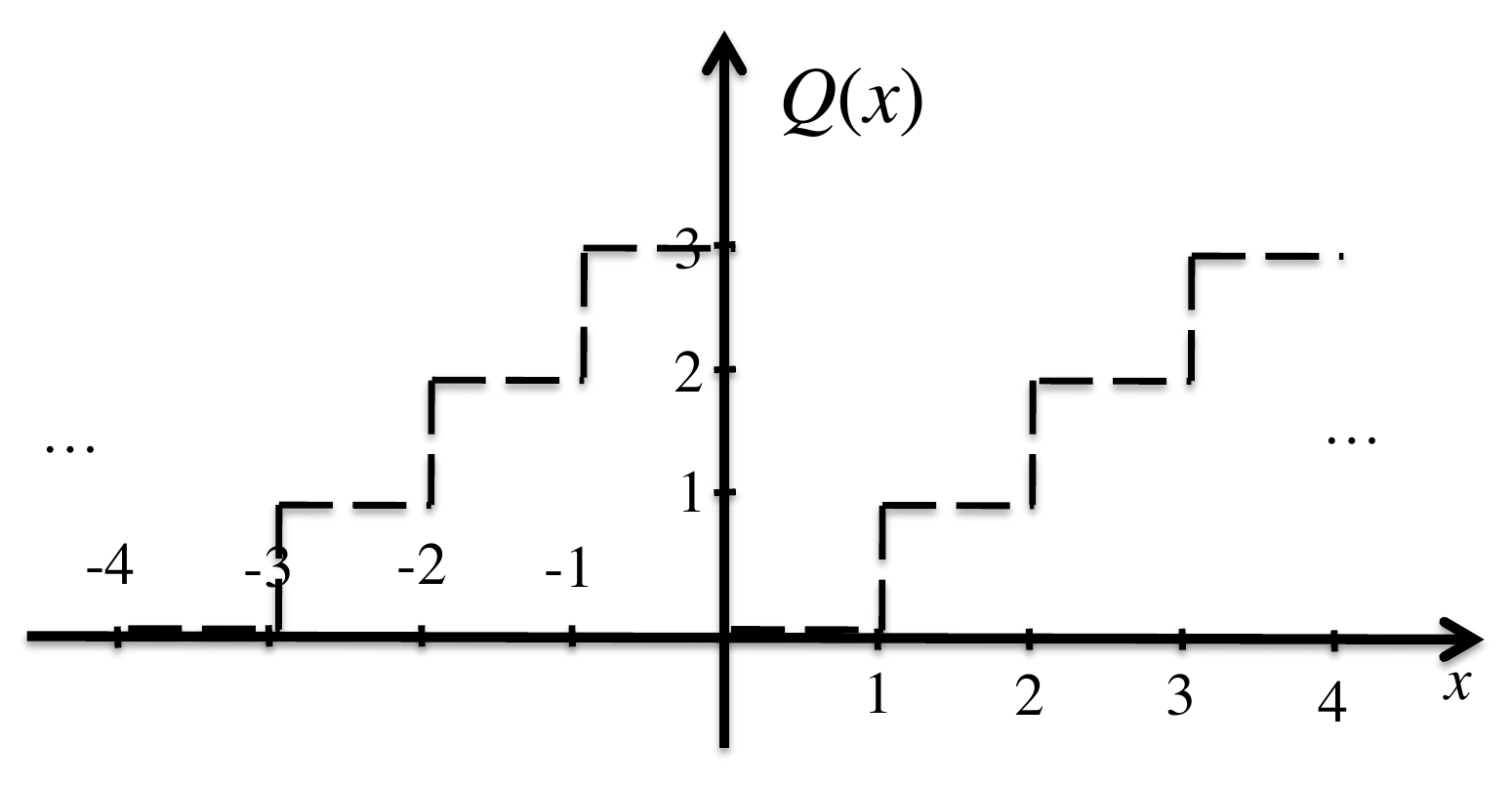}
  \end{minipage}
\end{minipage}
\hfill
\begin{minipage}[b]{.65\linewidth}
  \begin{center}
    \small{(b)}~~
    \includegraphics[width=.9\linewidth]{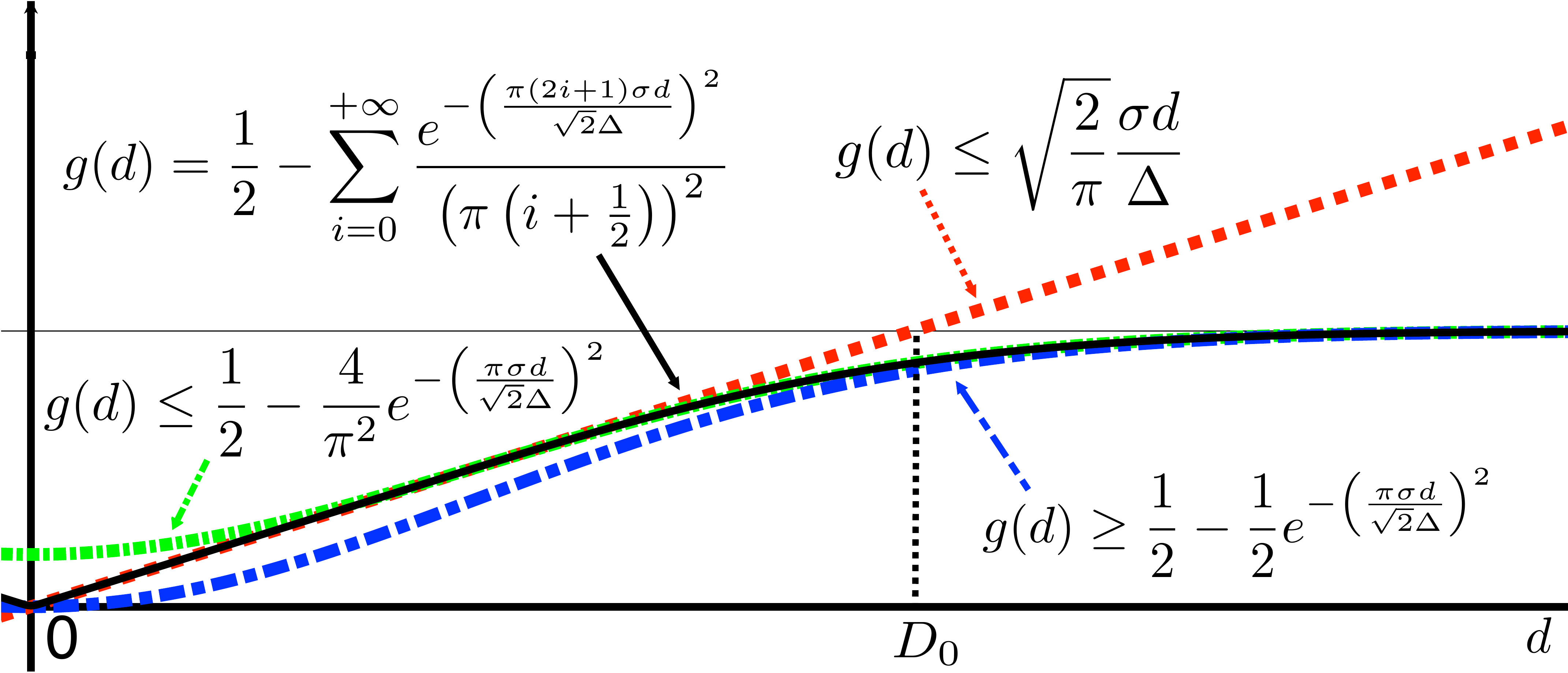}
  \end{center}
\end{minipage}
\\

\noindent \begin{minipage}[b]{\linewidth} \small{(c)}\hfill
  \includegraphics[width=.45\linewidth]{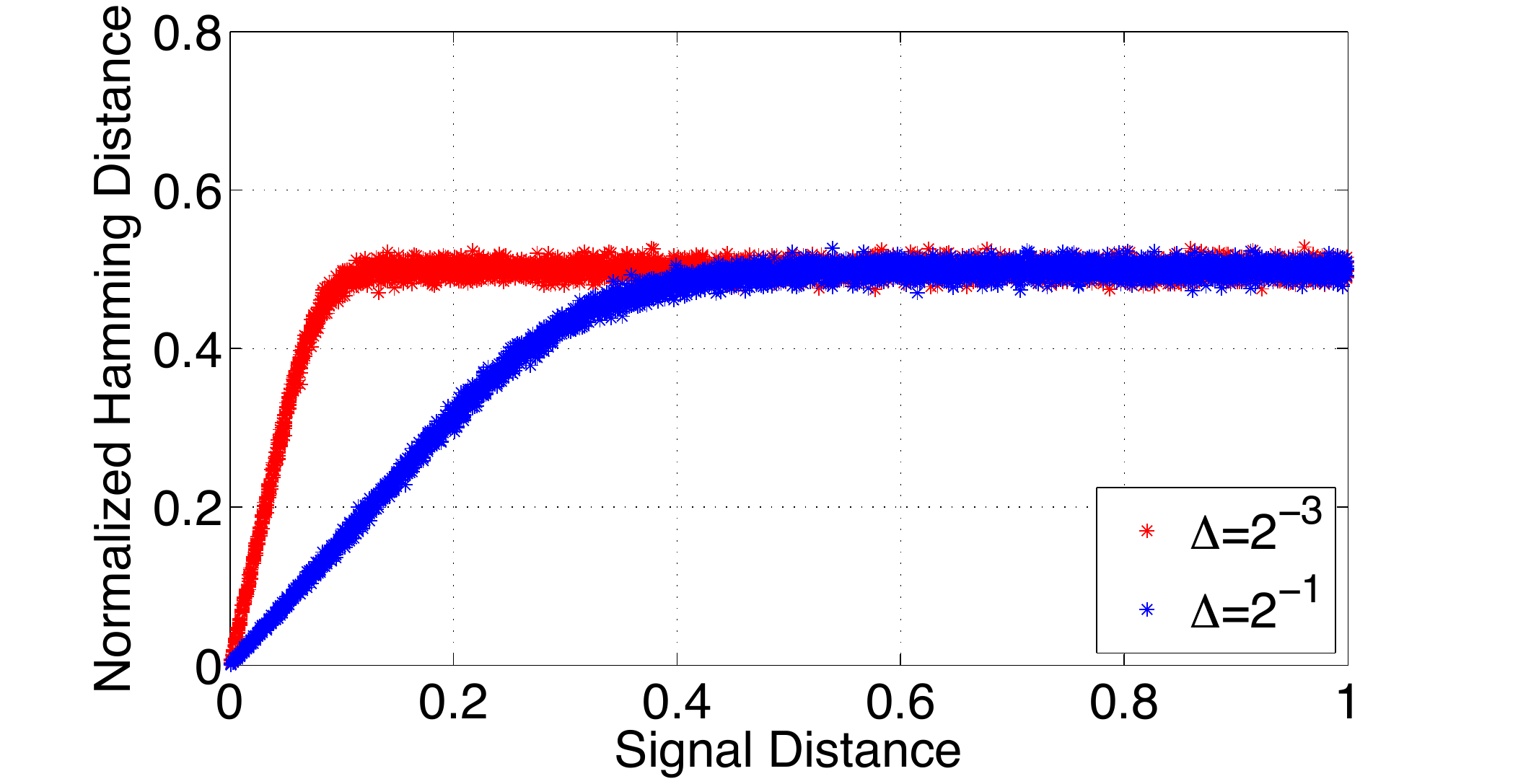}\hfill
  \includegraphics[width=.45\linewidth]{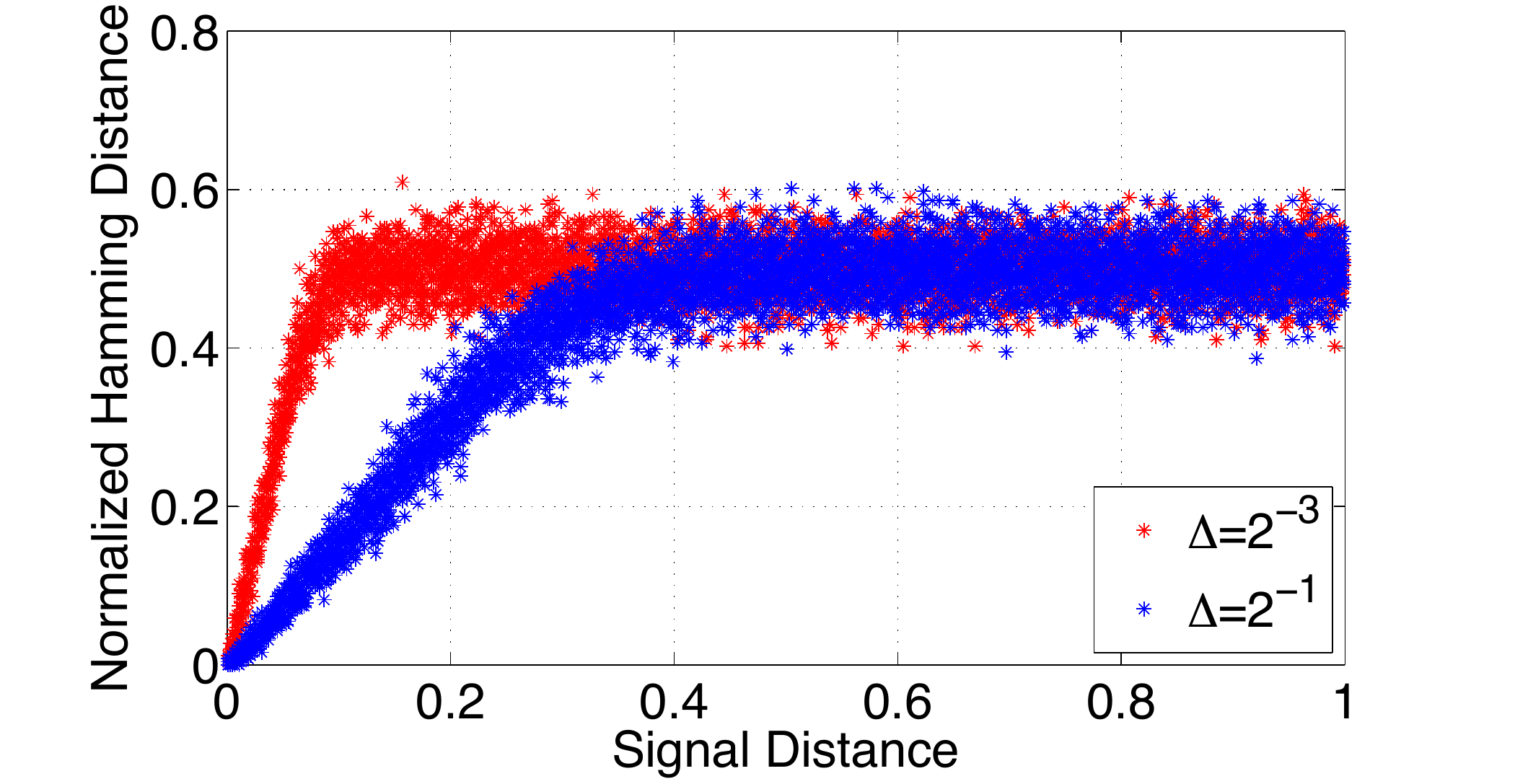}
      \end{minipage}

  \caption{(a) This non-monotonic quantization function $Q(\cdot)$
    allows for universal rate-efficient scalar quantization. This
    function is equivalent to using a classical multibit scalar
    quantizer, and preserving only the least significant bits while
    discarding all other bits. 1-bit shown on top, multi-bit shown on
    bottom (b) The embedding map $g(d)$ and its bounds produced by the
    1-bit quantization function in (a). (c) Experimental verification
    of the embedding for small and large $\Delta$ in high (left) and
    low (right) bitrates.}
  \label{fig:quant}
\end{figure*}

The breakthrough feature in this method is the modified $B$-bit scalar
quantizer, designed to be a periodic binary function with
non-contiguous quantization intervals, as shown in
Fig.~\ref{fig:quant}(a) for $B=1$ (top) and $B=2$ (bottom). The
quantizer can be thought of as a regular uniform quantizer, computing
a multi-bit representation of a signal and preserving only the least
significant bits (LSB) of the representation. For example, for a 1-bit
quantizer, scalar values in $[2l,2l+1)$ quantize to 1 and scalar
values in $[2l+1, 2(l+1))$, for any integer $l$, quantize to 0. If
$Q(\cdot)$ is a 1-bit quantizer, this method encodes using as many
bits as the rows of \mA, i.e., $M$ bits, and does not require
subsequent entropy coding. 

A large part of the development in this paper is inspired by (and
generalizes) the periodicity of the quantization function in binary
(1-bit) universal quantization. A multi-bit generalization of the
universal quantizer is shown in the bottom of Fig.~\ref{fig:quant}(a)
but has not, to-date, been analyzed or explored.

As discussed in~\cite{B_TIT_12}, the modified binary quantizer enables
efficient universal encoding of signals. Furthermore, this
quantization method is also an embedding of finite
signal sets~\cite{BR_WIFS11}. Specifically, on a set \sS\ with $L$ points, for all
$\vx,\vy\in\sS$, the embedding satisfies
\begin{align}
  g\left(\left\| \vx-\vy\right\|_2\right)-\epsilon\le
  d_H\left(f(\vx),f(\vy)\right)\le g\left(\left\|
  \vx-\vy\right\|_2\right)+\epsilon,
  \label{eq:universal_embedding}
\end{align}
with probability $1-2e^{2\log L-2\epsilon^2M}$ with respect to the
measure of \mA\ and \vw. In~\eqref{eq:universal_embedding},
$d_H(\cdot,\cdot)$ is the Hamming distance of the embedded signals,
the function $f(\cdot)$ is as specified in (\ref{eq:universal_quantization}), and
$g(d)$ is the map
\begin{align}
  g(d) =
  \frac{1}{2}-\sum_{i=0}^{+\infty}\frac{e^{-\left(\frac{\pi(2i+1)\sigma
        d} {\sqrt{2}\Delta}\right)^2}}{\left(\pi(i+1/2)\right)^2},
  \label{eq:ue_dist_map}
\end{align}
which can be bounded using, 
\begin{align}
 g(d)&\ge
  \frac{1}{2}-\frac{1}{2}e^{-\left(\frac{\pi\sigma
      d}{\sqrt{2}\Delta}\right)^2},\label{eq:bound_lower}\\
 g(d)&\le  \frac{1}{2}-\frac{4}{\pi^2}e^{-\left(\frac{\pi\sigma
      d}{\sqrt{2}\Delta}\right)^2},\label{eq:bound_upper_exp}\\
 g(d)&\le \sqrt{\frac{2}{\pi}}\frac{\sigma d}{\Delta},\label{eq:bound_upper_linear}
\end{align}
as shown in Fig.~\ref{fig:quant}(b). The map is approximately linear
for small $d$ and becomes a constant equal to $1/2$ exponentially fast
for large $d$, greater than a distance threshold $D_0$. The slope of
the linear section and the distance threshold $D_0$ is determined by
the parameter ratio $\sigma/\Delta$. In other words, the embedding
ensures that the Hamming distance of the embedded signals is
approximately proportional to the $\ell_2$ distance between the original signals,
as long as that $\ell_2$ distance was smaller than $D_0$. Note that a
piecewise linear function with slope
$\sqrt{\frac{2}{\pi}}\frac{\sigma}{\Delta}$ until $d=D_0$ and slope
equal to zero after that is a very good approximation
to~\eqref{eq:ue_dist_map}, in addition to being an upper bound.

To obtain a fixed bound on probability of failure, the additive
ambiguity $\epsilon$ in~\eqref{eq:universal_embedding} scales as
$\epsilon\propto 1/\sqrt{M}$, similar to the constant $\delta$ in the
multiplicative $(1\pm\delta)$ factor in J-L embeddings. It should be
noted, however, that universal embeddings use 1 bit per projection
dimension, for a total rate of $R=M$. The trade-off between $B$ and
$M$ under constant $R$ exhibited by quantized J-L embeddings does not
exist under 1-bit universal embeddings. Still, there is a performance
trade-off, controlled by the choice of $\Delta$
in~\eqref{eq:universal_quantization}, which is explored
in~\cite{BR_DCC13} and discussed in subsequent sections.

Figure~\ref{fig:quant}(c) demonstrates experimentally and provides
intuition on how the embedding behaves for smaller (red) and larger
(blue) $\Delta$ and for higher (left) and lower (right) bitrates. The
figure plots the embedding (Hamming) distance as a function of the
signal distance for randomly generated pairs of signals. The thickness
of the curve is quantified by $\epsilon$, whereas the slope of the
upward sloping part is quantified by $\Delta$.
 
Although universal embeddings perform very well in practice, a few
theoretical issues have not been resolved to date. One of the most
interesting questions is the extension to a multi-bit quantizer. While
such an extension, using the quantizer in the bottom of
Fig.~\ref{fig:quant}(a) is described in~\cite{B_TIT_12}, it has not
been analyzed and guarantees have not been provided. The techniques
used to provide the universal embeddings guarantee can fail or become
very tedious for a multi-bit analysis. 

Furthermore, the embedding guarantee has been shown
in~\cite{BR_WIFS11} to hold for finite point clouds and not for
infinite sets, such as sparse signals or manifolds. In~\cite{B_TIT_12}
it was shown that a different guarantee, on the distance of signals
that map to exactly the same binary vector, can be provided for such
sets. However, a general embedding guarantee, similar to the
extensions of the Johnson-Lindenstrauss lemma to the
RIP~\cite{BarDavDeV::2008::A-Simple-Proof} and to
manifolds~\cite{baraniuk2009random}, does not yet exist. In a number
of applications, such a guarantee is often desirable, if not necessary.

The development we present in the remainder of this paper addresses
both of these issues. Specifically, Sec.~\ref{sec:prob_construction}
provides a general description of how concentration of measure
inequalities are typically used to establish the embedding guarantees
and how these guarantees can be extended to hold for fairly general
embedding designs in infinite sets. It also examines the effect of
quantization on the embedding
guarantee. Section~\ref{sec:design_analysis} describes a general
embedding design approach which, combined with the development in
Sec.~\ref{sec:prob_construction} provides the desired guarantees for
multi-bit universal embeddings, of which binary universal embeddings become a
special case. The details are described in Sec.~\ref{sec:examples},
which uses universal embeddings as an example application of the
general theory.

Although not immediately relevant to this work, an
information-theoretic argument guarantees that binary universal
embeddings can preserve the query's privacy~\cite{BR_WIFS11}. These
are not explored in this work for the more general case. Although we
are confident that such guarantees can be provided, at least for
multi-bit universal embeddings, we defer such a development to a
separate publication.

%% file: bg_continuity.tex
\subsection{Lipschitz Continuity}
\label{sec:bg_continuity}
A very useful tool in the subsequent development is Lipschitz
Continuity. Lipschitz continuity enables us to bound how abruptly a
function may vary as its input varies. 
\begin{definition}
  A function $f:\sS\rightarrow\sW$ is Lipschitz-continuous, with Lipschitz constant
  $K$, if, for all $\vx,\vy\in\sS$:
  \begin{align}
    d_{\sW}(f(\vx),f(\vy))\le Kd_{\sS}(\vx,\vy)
  \end{align}
\end{definition}

While a number of functions we consider in the remainder of this paper
are Lipschitz continuous, some very interesting ones are not. For
example, quantization functions such as the ones in
Fig.~\ref{fig:quant}(a), are not Lipschitz continuous. Still, they do
exhibit piece-wise continuity properties that we should be able to
exploit and characterize. To that end, we introduce a generalization
of Lipschitz continuity, which we term \emph{$T$-part Lipschitz continuity}.

To characterize $T$-part Lipschitz continuity, we assume that the
function $f(\cdot)$ operates in a compact set \sS\ that can be
partitioned to $T$ subsets, such that $f(\cdot)$ is Lipschitz
continuous when its domain is restricted to each of the $T$ subsets.

\begin{definition}
  A function $f:\sS\rightarrow\sW$, where \sS\ is a compact set, is
  $T$-part Lipschitz continuous over \sS\ with Lipschitz constant $K$
  if there exists a finite partition of \sS\ to $T$ sets $\sS_t$,
  $t=1,\ldots,T$ such that for all $t$ and for all pairs $\vx,\vy$ in
  $\sS_t$ the Lipschitz property holds: $d(f(\vx,\vy))\le
  Kd(\vx,\vy)$. 
\end{definition}

\begin{definition}
  A function $f:\sS\rightarrow\sW$, where \sS\ is a compact set, is
  $T$-part constant if it is $T$-part Lipschitz continuous with $K=0$.
\end{definition}

Note that $T$-part Lipschitz continuity is a much more permissive
condition compared to piece-wise continuity, as typically understood
and defined. For example, the indicator function of rational numbers
\begin{align}
  I_{\mathbb{Q}}(x)=\left\{
  \begin{array}{rl}
    1,&~\mathrm{if}~x~\mathrm{is~rational}\\
    0,&~\mathrm{if}~x~\mathrm{is~irrational}\\
  \end{array}
  \right.
\end{align}
is nowhere continuous and definitely not Lipschitz
continuous. However, it is $2$-part constant, as we can split its 
domain, $\Reals$ to two sets $S_1=\mathbb{Q}$ and
$S_2=\Reals\setminus\mathbb{Q}$ over which $I_{\mathbb{Q}}(x)$ is
constant, i.e., $K=0$. Of course, the universal quantization functions
in Fig.~\ref{fig:quant}(a) are $2^B$-part Lipschitz constant, where
$B$ is the number of bits available to represent the quantization bins.

A function that is piece-wise continuous with a finite number of
pieces, $T$, is also $T$-part Lipschitz continuous. However,
piece-wise continuous functions with an infinite number of pieces may
not fit our definition. A conventional infinite uniform quantizer, for
example, is not $T$-part Lipschitz continuous. If, instead, the
quantizer has positive and negative saturation points, then there is a
finite number of quantization intervals, and the function becomes
$T$-part constant, with $T$ equal to the number of quantization
levels. Typically $T=2^B$, where $B$ is the number of quantization
bits.

We should note that $1$-part Lipschitz continuity coincides with the
traditional definition of Lipschitz continuity. Furthermore, a
$T$-part Lipschitz continuous function is also $T+1$-part Lipschitz
with the same constant $K$. We use the term ``{\em exactly} $T$-part
Lipschitz'', when $T$ represents the minimum number of partitions that can be
used to satisfy the definition of $T$-part Lipschitz continuity.

\begin{definition}
  A function $f:\sS\rightarrow\sW$, where \sS\ is a compact set, is
  {\em exactly} $T$-part Lipschitz continuous over \sS\ with Lipschitz
  constant $K$ if it is $T$-part Lipschitz continuous with that
  constant but is not $(T-1)$-part Lipschitz continuous with the same
  constant.
\end{definition}

%% file: probconstruct.tex
\section{Probabilistic Embedding Construction}
\label{sec:prob_construction}
As mentioned above, most embedding literature relies on randomized
constructions. Typically, such embeddings are demonstrated on finite
sets and often extended to cover infinite sets. However the extension
is often not trivial. Furthermore, the tools developed in the
literature,
e.g.,~\cite{BarDavDeV::2008::A-Simple-Proof,baraniuk2009random,raginsky2009locality,JLBB_1bit_2011},
are typically specific to each embedding design. Departing from this embedding-specific approach, we now
develop general methods that will allow us to extend a wide variety of embedding
designs from point clouds to infinite signal sets.

\input{pc_defandconcentration}
\input{pc_infinitecontinuous}

\input{pc_infinitediscontinuous}

\input{pc_quantized}

%% file: pc_defandconcentration.tex
\subsection{Embeddings and Concentration of Measure}
\label{sec:embedd-conc-meas}
Our goal in this paper is to present a fairly general framework to
design and analyze embeddings that approximately preserve distances
between signals. 
Typically, such embeddings are designed in a probabilistic manner by drawing
the embedding function $f$ randomly from a family of functions. For
example, in various constructions of Johnson-Lindenstrauss embeddings
or the Restricted Isometry Property (RIP) the function is a linear map
$f(\vx)=\mA\vx$, with the elements of $\mA$ drawn randomly from a
variety of possible
distributions~\cite{achlioptas03css,dasgupta03rsa,cande2008introduction,BarDavDeV::2008::A-Simple-Proof,CandesRIP}. More
recently, in quantized and phase embeddings, the linear map is followed
by a quantization or phase-extraction
operation~\cite{JLBB_1bit_2011,LRB_MMSP12,B_SampTA13,B_SPARS13,B_SPIE2013_APQPE,BR_WIFS11,BM_SAMPTA2015,plan2013robust,plan2013one,plan2014dimension,Jacques15buffon,jacques2015small}.

Since the embedding function is randomized, we can only prove that the
embedding is a $(g,\delta,\epsilon)$ embedding with high
probability. Most proofs rely on concentration of measure arguments to
show that~\eqref{eq:general_embedding} holds on a pair of points
$\vx,\vy\in\sS$ with high probability. Typically, the failure
probability decays exponentially with the number of measurements,
i.e., with the dimensionality of the embedding space $M=\dim(\sW)$. In
other words, the embedding fails on a pair of points with probability
bounded by $\Omega(e^{-Mw(\delta,\epsilon)})$, where
$w(\delta,\epsilon)$ is an increasing function of $\epsilon$ and
$\delta$ that quantifies the concentration of measure exhibited by the
randomized construction.

Once the embedding guarantee is established for a pair of signals, a
union bound can be used to extend it to a finite set of signals. If
the set \sS\ is finite, containing $Q$ points, then the probability
that the embedding fails is upper bounded by
$\Omega(Q^2e^{-Mw(\delta,\epsilon)})=\Omega(e^{2\log
  Q-Mw(\delta,\epsilon)})$, which decreases exponentially with $M$, as
long as $M=O(\log(Q))$.

Unfortunately this union bounding approach does not work for infinite sets, such as
signal spaces or sparse signals. Instead, to
establish~\eqref{eq:general_embedding} for such sets, a covering of the
set is constructed using a finite number of $\epsilon$-balls. The
concentration of measure is established for the centers of balls and
is then extended to all points of the balls using the continuity
properties of the embedding map. For
example,~\cite{BarDavDeV::2008::A-Simple-Proof} exploits the
properties of the distance metric to establish the RIP for all
$K$-sparse signals. This approach can be used if the distance map is
the identity, $g(d)=d$, but does not generalize very well. In the next
section, we describe a more general approach that can be used for
arbitrary Lipschitz-continuous distance and embedding maps.

%% file: pc_infinitecontinuous.tex
\subsection{Embedding of Infinite Sets Using Continuous Maps}
\label{sec:infinite_continuous}
To exploit Lipschitz continuity, we start with the randomized
embeddings as described in the previous section, i.e., for which we
can show that given a pair of points $\vx,\vy\in\sS$, the embedding
guarantee~\eqref{eq:general_embedding} holds with probability greater
than $1-ce^{-Mw(\delta,\epsilon)}$. We also assume the embedding map
$g(d)$ is Lipschitz-continuous with constant $K_g$ and that the
embedding map $f(\cdot)$ is Lipschitz-continuous with constant
$K_f$. Next, we use $C_{\varepsilon}^{\sS}$ to denote the covering
number of the signal set \sS, i.e., the smallest number of points
$\vq\in\sS$ s.t. for all $\vx\in\sS$, $\inf_{\vq}\|\vx-\vq\|\le
\varepsilon$. Its logarithm, $E_{\varepsilon}^{\sS}=\log
C_{\varepsilon}^{\sS}$ is the Kolmogorov $\varepsilon$-entropy of the
set. Appendix~\ref{app:proof_continuous} proves the following.

\begin{theorem}
  Consider a signal set \sS\ with $r$-covering number $C_{r}^{\sS}$
  and a $(g,\delta,\epsilon)$ probabilistic embedding to an
  $M$-dimensional space that fails with probability smaller than
  $ce^{-Mw(\delta,\epsilon)}$ on a pair of points. If the embedding
  map $f(\cdot)$ is $K_f$-Lipschitz continuous and the distance map
  $g(\cdot)$ is $K_g$-Lipschitz continuous, then for some $\alpha > 0$ the embedding is a
  $(g,\delta,\epsilon+\alpha))$ embedding that holds with probability
  greater than $1-ce^{2E_r^{\sS}-Mw(\delta,\epsilon)}$
  on all pairs of signals $\vx,\vy$ in \sS, where
  $r=\frac{\alpha}{(1+\delta)2K_g+2K_f}$.
  \label{thm:emb_continuous}
\end{theorem}

As typical with such proofs, the constants are difficult to pin down
accurately. However, the main takeaway is that the embeddings preserve
distances of an infinite set as long as the number of measurements $M$
is in the order of the Kolmogorov entropy of the set for a radius that
depends on the desired accuracy.

We should note that this theorem introduces an additive ambiguity
$\alpha$, even if the original embedding has $\epsilon=0$. For general
embedding maps, we do not believe this additive ambiguity can be
eliminated. In the special case of linear embedding functions
$f(\cdot)$ and linear distance maps $g(\cdot)$ the additive constant
can be eliminated using proof techniques such as the ones in,
e.g.,~\cite{BarDavDeV::2008::A-Simple-Proof,blumensath2009sampling}.

%% file: pc_infinitediscontinuous.tex
\subsection{Embedding of Infinite Sets Using Discontinuous Maps}
\label{sec:infinite_discontinuous}
To extend the mapping to discontinuous embeddings we separate the
randomized embedding $f(\cdot)$ into its components
$f_m(\cdot),~m=1,\ldots,M$, and examine how the embedding behaves on
balls \ball{r/2}{\vx}\ of diameter $r$, i.e., radius $r/2$. In
particular, we first examine the behavior of each function component
$f_m(\cdot)$ with domain restricted to a given ball
\ball{r/2}{\vx}\ with respect to its $T$-part Lipschitz
continuity. 

Assuming that each function $f_m:\ball{r/2}{\vx}\rightarrow\Reals$ is
exactly $T_m$-part Lipschitz over the ball, then we can partition the
ball into $T_m$ sets $S_{t_m},$ $t_m=1,\ldots,T_m$, over which the
$f_m(\cdot)$ is Lipschitz continuous. We can then define
$S_{t_1,\ldots,t_M}=S_{t_1}\cap\ldots\cap S_{t_M}$, which is a
partition of \ball{r/2}{\vx}\ of $T_1\times\ldots\times T_M$ sets such
that all $f_m$ are Lipschitz continuous over each
$S_{t_1,\ldots,t_M}$.

Next, we need to quantify $T_m$ for each $m$. Since the embedding is
randomized, we use $P_T$ to denote the probability that $f_m(\cdot)$
is exactly $T$-part Lipschitz over a ball \ball{r/2}{\vx}. We assume
that the probability that $f_m(\cdot)$ is exactly $T$-part Lipschitz
over \ball{r/2}{\vx}\ is independent from the probability that
$f_{m'}(\cdot), m\ne m'$ is $T$-part Lipschitz over the same
\ball{r/2}{\vx}. This, for example, is ensured if $f_m(\cdot)$ is
drawn independently for each $m$, as is typically the case. We assume
that the probability $P_T$ is a decreasing function of the ball radius
$r$ itself but does not depend on the ball in any other way. We also
select a maximum $T_{\mathrm{max}}$ beyond which the probability that
a function $f_m(\cdot)$ is exactly $T$-part Lipschitz continuous, is
negligible or zero.

\begin{theorem}
  Consider a signal set \sS\ with $r$-covering number $C_{r}^{\sS}$
  and a $(g,\delta,\epsilon)$ probabilistic embedding to an
  $M$-dimensional space that fails with probability smaller than
  $ce^{-Mw(\delta,\epsilon)}$ on a pair of points. If each coordinate
  of the embedding map $f_m(\cdot)$ is exactly $T$-part Lipschitz
  continuous with probability less than $P_T$, as described above, and
  the distance map $g(\cdot)$ is $K_g$-Lipschitz continuous, then the
  embedding is a $(g,\delta,\epsilon+\alpha)$ embedding that with
  probability greater than
  $1-(ce^{2E_{r/2}^\sS+c_1M-Mw(\delta,\epsilon)}+
  T_{\mathrm{max}}e^{-2c_0^2M}+P_F)$ holds on
  all pairs of signals $\vx,\vy$ in \sS, for any $T_{\mathrm{max}}$,
  where $r=\frac{\alpha}{(1+\delta)2K_g+2K_f}$,
  $P_F= \sum_{T=T_{\mathrm{max}}+1}^\infty P_T$,
  $c_1=\sum_{T=2}^{T_{\mathrm{max}}}P_T(1+c_0)\log T$, and any
  $\alpha<1$.
\label{thm:emb_discontinuous}
\end{theorem}

The proof details can be found in
Appendix~\ref{app:proof_discontinuous}.

Assuming $P_F$ is negligible, the embedding fails with probability at
most $ce^{2E_{r/2}^\sS+c_1M-Mw(\delta,\widetilde{\epsilon}-\alpha)}
+T_{\mathrm{max}}e^{-2c_0^2M}+P_F$, which decays exponentially with
$M$ as long as $c_1<w(\delta,\epsilon)$ and $M=O(E_{r/2}^\sS)$. The
inequality holds for sufficiently small $r$ since $c_1$ decreases with
$r$ for most randomized embedding constructions.

%% file: pc_quantized.tex
\subsection{Quantized Embeddings}
\label{sec:quantized-embeddings}
Although quantization of the embedding can be analyzed using the
framework we describe above, it is often more convenient, especially
in the case of high-rate quantization to consider it separately, as an
additional step after the projection.

We examine a $(g,\delta,\epsilon)$ embedding which is subsequently
quantized using an $M$-dimensional vector quantizer $Q(\cdot)$. We
assume the quantization error is bounded, i.e., $d(Q(\vx),\vx)\le
E_Q$. The triangle inequality,
$|d_{\sW}(f(\vx),f(\vw))-d_{\sW}(Q(f(\vx)),Q(f(\vw)))|\le 2E_Q$, 
implies that the quantized embedding guarantee becomes a
$(g,\delta,\epsilon+2E_Q)$ embedding, with guarantee
\begin{align}
  (1-\delta)g\left(d_{\sS}(\vx,\vy)\right)-\epsilon-2E_Q\le
  d_{\sW}\left(Q(f(\vx)),Q(f(\vy))\right)\le
  (1+\delta)g\left(d_{\sS}(\vx,\vy)\right)+\epsilon+2E_Q.
  \label{eq:quantized_embedding}
\end{align}

\begin{theorem}
  Consider a $(g,\delta,\epsilon)$ embedding $f(\cdot)$ and a
  quantizer $Q(\cdot)$ with worst case quantization error $E_Q$, then
  the quantized embedding, $Q(f(\cdot))$, is a
  $(g,\delta,\epsilon+2E_Q)$ embedding.
  \label{thm:quantized_embedding}
\end{theorem}

In the specific case of a uniform scalar quantizer with quantization
interval $\Delta$, the $M$-dimensional quantization $\ell_2$ error is
bounded by $E_Q\le\sqrt{M}\Delta/2$, assuming the quantizer is
designed such that it does not saturate or such that the saturation
error is negligible. The interval of the quantizer is a function of
the number of bits $B$ used per coefficient $\Delta=2^{-B+1}S$, where
$S$ is the saturation level of the quantizer. Given a fixed rate to be
used by the embedding, $R=MB$, the guarantee becomes
\begin{align}
  (1-\delta)g\left(d_{\sS}(\vx,\vy)\right)-\epsilon-2^{-\frac{R}{M}+1}\sqrt{M}S\le
  \|Q(f(\vx))-Q(f(\vy))\|_2\le
  (1+\delta)g\left(d_{\sS}(\vx,\vy)\right)+\epsilon+2^{-\frac{R}{M}+1}\sqrt{M}S.
  \label{eq:su_quantized_embedding}
\end{align}
Note that the $\sqrt{M}$ factor can often be removed, depending on the
normalization of the embedding.

Of course, $\ell_2$ is not always the appropriate fidelity metric. If
the $d_\sS(\cdot,\cdot)$ corresponds to the $\ell_1$ distance, the
quantization error is bounded by $E_Q\le M\Delta/2$. Again, with care
in the normalization the $M$ factor can be removed. If, instead, the
$\ell_\infty$ norm is desired, the quantization error is bounded by
$E_Q\le\Delta/2$. 

One of the issues to consider in designing quantized embeddings using
a uniform scalar quantizer is the trade-off between the number of bits
per dimension and the total number of dimensions used. Since $R=MB$,
increasing the number of bits per dimension $B$ under a fixed bit budget
$R$, requires decreasing the number of dimensions $M$. While the
former reduces the error due to quantization, the latter will
typically increase the uncertainty in the embedding by increasing
$\delta$ and $\epsilon$.  

In the case of randomized embeddings, this trade-off can be quantified
through the function $w(\epsilon,\delta)$. Given a fixed probability
lower bound to guarantee the embedding, then
$M=\Omega(1/w(\epsilon,\delta))$. Since $w(\cdot,\cdot)$ is an
increasing function of $\epsilon$ and $\delta$, which quantify the
ambiguity of the embedding, reducing $M$ increases this ambiguity. On
the other hand, the quantization ambiguity, quantified in
$2^{-R/M+2}S\sqrt{M}$ decreases with $M$. This trade-off is explored,
for example, in the context of quantized J-L embeddings
in~\cite{LRB_MMSP12,RBV_SPIE13_Embeddings}. Although we describe the
trade-off for uniform scalar quantizers, the same issue exists for
non-uniform quantizers and for vector quantizers, manifested with
different constants but with the same order of magnitude effects
(e.g., see~\cite{jacques2013stabilizing}).

Sec.~\ref{sec:examples} provides examples of quantized embeddings,
examining cases where the quantization is analyzed as described here,
i.e., as an additional step after the embedding is performed, as well
as cases in which quantization is analyzed as part of the embedding
through the mechanism of $T$-part Lipschitz functions.

%% file: designandanalysis.tex
\section{Embedding Design and Performance Analysis}
\label{sec:design_analysis}

\input{da_design}
\input{da_matrixdesign}
\input{da_properties}
\input{da_analysis}

\input{da_kernelembeddings}

%% file: da_design.tex
\subsection{Embedding Design}
\label{sec:design}
Having provided all the necessary tools to establish the embedding
properties over a signal set, we next consider a fairly general
embedding design. Specifically, we consider the mapping
$\vy=h(\mA\vx+\vw)$, where the rows $\va_i$ of $\mA$ are randomly
chosen from some i.i.d. vector distribution and the elements of $\vw$
are chosen from an i.i.d. distribution uniform in $[0,1)$. We denote
  the projection through \mA\ using $\vu=\mA\vx$. 

We restrict our attention to functions $h_g(t)$ with finite support,
restricted in $[0,1)$ without loss of generality, and their periodic
  extension $h(t)$ with period 1, i.e., such that $h(t)=h_g(t)$ for
  $t\in [0,1)$ and $h(t)=h(t+1)$. We use $H_g(\xi)$ and $H(\xi)$ to
    denote their respective Fourier transform, $R_{h_g}(\tau)$ and
    $R_{h}(\tau)$ to denote their deterministic autocorrelations, and
    $P_{h_g}(\xi)$ and $P_h(\xi)$ to denote their power
    spectrum. Their relationship is summarized below.
\begin{align}
  h_g(t)=0~\mathrm{if}~t\notin[0,1)~&\overset{\mathcal{F}}{\longleftrightarrow}~H_g(\xi)\\
  h(t)=\sum_{k=-\infty}^{+\infty}h_g(t-k)\delta(t-k)~&\overset{\mathcal{F}}{\longleftrightarrow}~H(\xi)=H_g(\xi)\sum_{k=-\infty}^{+\infty}\delta(\xi-k)=\sum_{k=-\infty}^{+\infty}H_k\delta(\xi-k)\\
  R_{h_g}(\tau)=\int_{-\infty}^{+\infty}h_g(t)h_g(t-\tau)dt~&\overset{\mathcal{F}}{\longleftrightarrow}~P_{h_g}(\xi)=|H(\xi)|^2\\
  R_h(\tau)=\int_{0}^{+1}h(t)h(t-\tau)dt~&\overset{\mathcal{F}}{\longleftrightarrow}~P_h(\xi)=|H_g(\xi)|^2\sum_{k=-\infty}^{+\infty}\delta(\xi-k)=\sum_{k=-\infty}^{+\infty}|H_k|^2\delta(\xi-k),
\end{align}
where $H_k=H_g(k)$ denotes the Fourier series coefficients of $h(t)$.
Note that, to avoid convergence issues, the autocorrelation of periodic
functions is defined as the integral over a single period, in contrast
to the finite-support autocorrelation defined as the integral over all
$\Reals$. Although we use the same notation for simplicity, the
appropriate use should be clear from the context.

We first examine the behavior of a single coefficient of \vy, i.e.,
$y=h(\langle \va,\vx\rangle+w)$, where \va\ is the corresponding row
of \mA\ and $w$ the corresponding coefficient of \vw. If we measure a
pair of signals \vx\ and $\vx'$ at distance $d=d_{\sS}(\vx-\vx')$
apart, their (signed) projected distance, denoted $l=\langle \va,
\vx-\vx' \rangle$, is a random variable with density conditioned on
$d$ denoted using $f_l(\cdot|d)$ and characteristic function denoted
using $\phi_l(\xi|d)$. Then, we can prove the following theorem.

\begin{theorem}
  \label{thm:point_embedding}
  Consider a set \sS\ of $Q$ points in $\Reals^N$, measured using
  $\vy=h(\mA\vx+\vw)$, with \mA, \vw, and $h(t)$ as above. With
  probability greater than
  $1-e^{2\log Q-2M\frac{\epsilon^2}{\bar{h}^4}}$ the following
  holds
  \begin{align}
    g(d)-\epsilon\le\frac{1}{M}\left\|\vy-\vy'\right\|_2^2\le
    g(d)+\epsilon
    \label{eq:thm_guarantee}
  \end{align}
  for all pairs $\vx,\vx'\in\sS$ and their corresponding measurements
  $\vy,\vy'$, where 
  \begin{align}
    g(d)=2\sum_k|H_k|^2(1-\phi_l(2\pi k|d)).
    \label{eq:dist_map}
  \end{align}
  defines the distance map of the embedding.
\end{theorem}
The proof is provided in Appendix~\ref{app:proof_pointembedding}. Note
that $\phi_l(0)=1$ for any distribution. Thus, the DC component $H_0$
of the embedding map does not affect the distance map
in~\eqref{eq:dist_map}.

To show that this is a $(g,0,\epsilon)$ embedding, we derive a bound
on the $\ell_2$ distance, instead of its square, which follows easily
from the fact that
$\sqrt{x\pm\epsilon}\lessgtr\sqrt{x}\pm\sqrt{\epsilon}$:
\begin{corollary}
  \label{cor:root_point_embedding}
  Consider the signal set \sS, defined and measured as in
  Thm.~\ref{thm:point_embedding}. With probability greater than
  $1-e^{2\log Q-2M\left(\frac{\epsilon}{\bar{h}}\right)^4}$ the following holds
  \begin{align}
    \widetilde{g}(d)-\epsilon\le\frac{1}{\sqrt{M}}\left\|\vy-\vy'\right\|_2\le \widetilde{g}(d)+\epsilon
  \end{align}
  for all pairs $\vx,\vx'\in\sS$ and their corresponding measurements
  $\vy,\vy'$, where $\widetilde{g}(d)=\sqrt{g(d)}$.
\end{corollary}
For $\epsilon<1$, we can derive a tighter bound using
$\sqrt{x\pm\epsilon}\lessgtr\sqrt{x}\pm\epsilon$:
\begin{corollary}
  \label{cor:root_point_embedding_tight}
  Consider the signal set \sS, defined and measured as in
  Thm.~\ref{thm:point_embedding}. With probability greater than
  $1-e^{2\log Q-2M\frac{\epsilon^2}{\bar{h}^4}}$, if $\epsilon\le 1$ the
  following holds
  \begin{align}
    \widetilde{g}(d)-\epsilon\le\frac{1}{\sqrt{M}}\left\|\vy-\vy'\right\|_2\le \widetilde{g}(d)+\epsilon
  \end{align}
  for all pairs $\vx,\vx'\in\sS$ and their corresponding measurements
  $\vy,\vy'$, where $\widetilde{g}(d)=\sqrt{g(d)}$.
\end{corollary}

In a similar manner, we can establish that
$E\{\|y\|_2^2\}=\sum_k|H_k|^2$ and, therefore,
\begin{align}
  \sum_k|H_k|^2-\epsilon\le\frac{1}{M}\|y\|_2^2\le\sum_k|H_k|^2+ \epsilon,
  \label{eq:norm_bound}
\end{align}
with probability greater than $1-2e^{\log
  Q-2M\frac{\epsilon^2}{\bar{h}}}$. The proof is similar to the proof
of Thm.~\ref{thm:point_embedding} in
App.~\ref{app:proof_pointembedding}, and we omit it here for
brevity. However, we should note that, to ensure
both~\eqref{eq:thm_guarantee} and~\eqref{eq:norm_bound} hold, the
union bound should be taken over both the point pairs and the points in \sS. Still, the probability that both
hold remains bounded by $1-Q^2e^{-2M\frac{\epsilon^2}{\bar{h}}}$
because of the bounding approximations used in
proving~Thm.~\ref{thm:point_embedding} in
App.~\ref{app:proof_pointembedding} to simplify the theorem
statement. We should also note that in certain cases, such as binary
embeddings, the value of $\|y\|_2^2$ can be exactly
computed~\cite{BM_SAMPTA2015}.

Of course, using the results of the
Section~\ref{sec:prob_construction} with
$w(\epsilon)=2(\epsilon/\bar{h})^4$ or
$w(\epsilon)=2\epsilon^2/\bar{h}^4$ we can trivially establish the
embedding over infinite sets.

%% file: da_matrixdesign.tex
\subsection{Projection Randomization and the Distance Map}
\label{sec:rand-choic}
There are several choices for the randomization of the embedding
projection matrix $\mA$, which result in interesting properties of the
embedding map and the distances preserved. Specifically, the distance
$d=d_{\sS}(\vx-\vx')$ of two signals affects $l$ according to the
characteristic function $\phi_l(\xi|d)$. By selecting the appropriate
distribution on $\mA$, the map can be designed to preserve a variety
of distances. We examine two particularly useful examples, preserving
maps of $\ell_2$ and $\ell_1$ distances. The subsequent discussion, as
well as the discussion in Sec.~\ref{sec:embedd-kern-inner}, exploits
and generalizes results in~\cite{Rahimi07,B_TIT_12}.

\subsubsection{Mapping $\ell_2$ distances} 
If elements of $\mA$ are chosen from an i.i.d. Normal distribution
with variance $\sigma^2$ then $l$ is a normally distributed random
variable with variance $(d_{\ell_2}\sigma)^2$, where
$d_{\ell_2}=\|\vx-\vx'\|_2$. In other words, it is natural to set the
distance $d_{\sS}(\cdot,\cdot)$ above as the $\ell_2$ distance. In
that case, the characteristic function is that of a normal
distribution $\phi_l(\xi|d)=\phi_{\mathcal{N}(0,\sigma^2
  d^2)}(\xi)=e^{-\frac{1}{2}(\sigma d\xi)^2}$, and the distance map
becomes
\begin{align}
  g(d)=2\sum_k|H_k|^2(1-e^{-2(\pi\sigma dk)^2}),
  \label{eq:gaussian_dist_map}
\end{align}
with $d$ measuring the $\ell_2$ distance.

\subsubsection{Mapping $\ell_1$ distances} 
If, instead, elements of $\mA$ are drawn from an i.i.d. 
Cauchy distribution with zero location parameter and scale parameter $\gamma$, i.e., density
\begin{align}
  f_a(x) = { 1 \over \pi \gamma } \left[ { \gamma^2 \over x^2 +
      \gamma^2  } \right],
\end{align}
with corresponding characteristic function
$\phi_a(\xi)=e^{-\gamma|\xi|}$, then $l$ is a sum of independent
Cauchy-distributed random variables. It is straightforward to show
that the resulting characteristic function is a function of the
$\ell_1$ distance of the two signals, $d_{\ell_1}=\|\vx-\vx'\|_1$. In
particular,
\begin{align}
  \phi_l(\xi|d_{\ell_1})=e^{-\gamma d_{\ell_1}|\xi|},
\end{align}
and the corresponding distance map becomes
\begin{align}
  g(d)=2\sum_k|H_k|^2(1-e^{-2\pi\gamma dk}),
\end{align}
with $d$ in this case measuring the $\ell_1$ distance. This enables
direct embedding of an $\ell_1$ space to an $\ell_2$ space, in
contrast to solutions that first map $\ell_1$ to a much larger
$\ell_2$ space through a ``unary'' expansion and then embed the
mapping, as done, for example,
in~\cite{linial1995geometry,RBV_SPIE13_Embeddings}.

%% file: da_properties.tex
\subsection{Properties of the distance map}
Although it would be desirable to be able to generate any distance map
desired, the properties of distance computation impose constraints on
the distance maps that are possible. In particular, within the
$\epsilon$ and $\delta$ error bounds of the embedding, the distance
map should be subadditive.
\begin{definition}
  A function $g(x)$ is $(\epsilon,\delta)$-subadditive for all $a,b$
  in its domain:
  $(1-\epsilon)g(a+b)-\delta\le g(a)+g(b)$ 
\end{definition}

For the case of the distance map $g(d)$ in a $(g,\delta,\epsilon)$
embedding, the following proposition is proven in
Appendix~\ref{sec:proof-subadditivity}:
\begin{proposition}
  \label{prop:subadditivity}
  Any distance map $g(\cdot)$ satisfying
  Def.~\ref{def:general_embedding} for a convex set \sS, is
  $(2\epsilon,3\delta)$-subadditive.
\end{proposition}

The subadditivity of the distance map imposes constraints on the
distance maps that are achievable with such a scheme. For example, a
distance map that is small for a range of distances cannot become
large immediately after; it is easy to show that for some positive $a$, if $g(d)<a$ for
$d<d_0$, then $g(d)<(2a+\delta)/(1-\epsilon)$ for $d\le 2 d_0$.

In addition to possessing the subadditivity property, distance maps $g(d)$ designed in this
section are comprised of linear combinations of increasing functions of $d$
bounded by $1$. Thus, they are bounded by
\begin{align}
  g(d)&\le\lim_{d\rightarrow\infty}g(d)=2\sum_k\left|H_k\right|^2=2\int_0^1|h(t)|^2dt=2R_h(0)
\end{align}
Since the square root is also a monotonic function, $\sqrt{g(d)}$,
used in Cor.~\ref{cor:root_point_embedding}, is also increasing and
bounded by $\sqrt{2R_h(0)}$. However, it remains an open question whether all
distance maps satisfying~\eqref{eq:general_embedding} should satisfy
some monotonicity constraint.

%% file: da_analysis.tex
\subsection{Error Analysis}
\label{sec:error-analysis}
To understand the performance of an embedding in distance computation
and to guide our design we want to understand how well the embedding
captures the distance. The main question is: given a distance
$d_{\sW}$ between two embedded signals in the embedding space \sW, how
confident are we about the corresponding distance between the original
signals in the signal space \sS? The function $g(\cdot)$ captures how
distance is mapped and can be inverted to approximately determine the
distance $d_{\sS}$ in the signal space. On the other hand, the
constants $\delta$ and $\epsilon$ capture the ambiguity in the
opposite direction, i.e., the ambiguity in the embedding space given
the distance in the signal space. Pictorially, taking
Fig.~\ref{fig:quant}(c) as an example, \eqref{eq:general_embedding}
characterizes the thickness of the curves taking a vertical slice of
the plots, while we are now interested in the thickness revealed by
taking a horizontal slice instead.

To capture the desired ambiguity, we can reformulate the embedding
guarantees as
\begin{align}
  g^{-1}\left(\frac{d_{\sW}\left(f(\vx),f(\vy)\right)-\epsilon}
  {(1+\delta)}\right) \le d_{\sS}(\vx,\vy)\le
  g^{-1}\left(\frac{d_{\sW}\left(f(\vx),f(\vy)\right)+\epsilon}
  {(1-\delta)}\right),
\end{align}
which for small $\delta$ and $\epsilon$ can be approximated using the
Taylor expansion of $1/(1\pm\delta)$:
\begin{align}
  g^{-1}\left(\left(d_{\sW}\left(f(\vx),f(\vy)\right)-\epsilon\right)
  \left(1-\delta\right)\right) \lesssim d_{\sS}(\vx,\vy)\lesssim
  g^{-1}\left(\left(d_{\sW}\left(f(\vx),f(\vy)\right)+\epsilon\right)
  \left(1+\delta\right)\right),
\end{align}
 Assuming that $g(\cdot)$ is differentiable, we can approximate the
 inequality using the Taylor expansion of $g^{-1}(\cdot)$ around
 $d_{\sW}\left(f(\vx),f(\vy)\right)$ and the fact that
 $(g^{-1})'(x)=1/g'(g^{-1}(x))$. Ignoring the second order term
 involving $\epsilon\cdot\delta$, and defining the signal distance
 estimate
 $\widetilde{d}_{\sS}=g^{-1}\left(d_{\sW}\left(f(\vx),f(\vy)\right)\right)$
 we obtain
 \begin{align}
   \widetilde{d}_{\sS} -\frac{\epsilon+\delta
     d_{\sW}\left(f(\vx),f(\vy)\right)}
             {g'\left(\widetilde{d}_{\sS}\right)}
             \lesssim d_{\sS}(\vx,\vy) \lesssim
             \widetilde{d}_{\sS}
             +\frac{\epsilon+\delta d_{\sW}\left(f(\vx),f(\vy)\right)}
             {g'\left(\widetilde{d}_{\sS}\right)}.
 \end{align}
In other words, given the distance $d_{\sS}$ between two signals in
the signal space and using $\widetilde{d}_{\sS}$ to denote the
estimate of this distance, the ambiguity is less than
\begin{align}
\left|d_{\sS}(\vx,\vy)-\widetilde{d}_{\sS}\right|
  \lesssim\frac{\epsilon+\delta d_{\sW}\left(f(\vx),f(\vy)
    \right)}{
    g'\left(\widetilde{d}_{\sS}\right)}.\label{eq:ambiguity}
\end{align}
Thus, ambiguity decreases by decreasing $\delta$ or $\epsilon$, or by
increasing the slope of the mapping.

%% file: da_kernelembeddings.tex
\subsection{Embeddings of Kernel Inner Products}
\label{sec:embedd-kern-inner}
Randomized projections, as first demonstrated
in~\cite{Rahimi07}, can be used to approximate kernel inner products
for some shift-invariant kernels. In addition to preserving distances,
the embeddings we describe also provide a more general kernel
approximation approach, generalizing the results in~\cite{Rahimi07}.

The inner product of the measurements $\langle \vy,\vy'\rangle$ can be
derived from the $\ell_2^2$ difference of the measurements,
$\|\vy-\vy'\|_2^2$. Specifically,
\begin{align}
  \|\vy-\vy'\|_2^2=\|\vy\|_2^2+\|\vy'\|_2^2-2\langle\vy,\vy'\rangle
  \Longrightarrow \langle\vy,\vy'\rangle =
  \frac{\|\vy\|_2^2+\|\vy'\|_2^2-\|\vy-\vy'\|_2^2}{2}.
  \label{eq:inner_expansion}
\end{align}
Thus, if $d_\sW(\vy,\vy')=\|\vy-\vy'\|_2^2$ in
Def.~\ref{def:general_embedding}, and
substituting~\eqref{eq:thm_guarantee} and~\eqref{eq:norm_bound}
in~\eqref{eq:inner_expansion}, we can show that the embedding can be
designed to approximate a kernel.
\begin{theorem}
  \label{thm:kernel_embedding}
  Consider a set \sS\ of $Q$ points in $\Reals^N$, measured using
  $\vy=h(\mA\vx+\vw)$, with \mA, \vw, and $h(t)$ as above. With
  probability greater than $1-e^{2\log Q-\frac{8}{9}M\frac{\epsilon^2}{\bar{h}^4}}$
  the following holds
  \begin{align}
    K(d)-\epsilon\le\frac{1}{M}\langle\vy,\vy'\rangle\le
    K(d)+\epsilon
    \label{eq:thm_kernel_guarantee}
  \end{align}
  for all pairs $\vx,\vx'\in\sS$ and their corresponding measurements
  $\vy,\vy'$, where 
  \begin{align}
    K(d)=\sum_k|H_k|^2\phi_l(k|d).
    \label{eq:kernel_map}
  \end{align}
  defines the kernel of the embedding.
\end{theorem}

%% file: examples.tex
\section{Embedding Examples}
\label{sec:examples}
\input{ex_qjl}

\input{ex_binaryuniversal}
\input{ex_multibituniversal}

%% file: ex_qjl.tex
\subsection{Quantized J-L Embeddings}
As a first example, we analyze quantized J-L embeddings. In classical
J-L embeddings, the distance map is the identity, i.e., $g(d)=d$, and
$\epsilon=0$. Starting with classical J-L embeddings, using the
development in Sec.~\ref{sec:quantized-embeddings}, we can derive the
quantized J-L embedding guarantees described in~\cite{LRB_MMSP12}. In
this case, the scaling of the embedding allows the removal of the
$\sqrt{M}$ term from~\eqref{eq:su_quantized_embedding}:
\begin{align}
  (1-\delta)\|\vx-\vy\|_2-2^{-\frac{R}{M}+1}S\le
  \|Q(f(\vx))-Q(f(\vy))\|_2\le
  (1+\delta)\|\vx-\vy\|_2+2^{-\frac{R}{M}+1}S.
\end{align}

Since $g(d)=d$, which has constant slope equal to 1, the denominator
in~\eqref{eq:ambiguity} is constant. To reduce the ambiguity, a system
designer should reduce the numerator as much as possible.  To do so,
as discussed in~\cite{LRB_MMSP12}, the designer confronts the
trade-off between the size of $\delta$ and $\epsilon$. The former is
controlled by the dimensionality of the projection, $K$, while the
latter by the bit-rate per dimension, $B$. The greater $K$ is, the
smaller $\delta$ is. Similarly, the greater $B$ is, the smaller
$\epsilon$ is.

As we mention above, the total bit-rate of the embedding is equal to
$R=KB$. In order to best use a given rate, the system designer should
explore the trade-off between fewer projection dimensions at more bits
per dimension and more projection dimensions at fewer bits per
dimension. This trade-off is explored in detail in~\cite{LRB_MMSP12},
where it is shown that, in the image retrieval application considered,
the best performance is achieved using $B=3$ or $4$ bits per dimension
and $K=R/3$ or $R/4$ dimensions, respectively. The performance of the
two choices is virtually indistinguishable and significantly better
than previous 1-bit approaches~\cite{CP,RP}, which use $B=1$, $R=K$.

%% file: ex_binaryuniversal.tex
\subsection{Binary Universal Embeddings}
\label{sec:uq_theory}
Universal Embeddings provide a more comprehensive example of the
development and analysis above. These embeddings are computed
using~\eqref{eq:universal_quantization} with the periodic quantizer
shown in Fig.~\ref{fig:quant}(a). With appropriate scaling on the
period of the quantizer, these embeddings satisfy exactly the
conditions of Thm.~\ref{thm:point_embedding}. 

\subsubsection{Embedding Map}
\label{sec:embedding-map}
The special case of binary universal embeddings has been extensively
studied in~\cite{B_TIT_12,BR_DCC13,BM_SAMPTA2015,BR_WIFS11}, providing
distance embedding results and kernel approximation guarantees. These
results, using the development in Sec.~\ref{sec:design_analysis},
become a special case of Thms.~\ref{thm:point_embedding}
and~\ref{thm:kernel_embedding}. Specifically, the quantizer $Q(x)$ in
Fig.~\ref{fig:quant}(a) has period $1$, i.e., $\widetilde{Q}(x)=Q(2x)$
has the correct period. The scaling can be incorporated in the
generation of $A$, i.e.,~\eqref{eq:universal_quantization} becomes
\begin{align}
  \label{eq:quantizer_scaling}
  Q\left(2
  \left(\frac{1}{2}\Delta^{-1}\mA\vx+\frac{1}{2}\Delta^{-1}\vw\right)\right)
  =\widetilde{Q}\left(\widetilde{\mA}\vx+\widetilde{\vw}\right),
\end{align}
where $\widetilde{\mA}$ is drawn i.i.d. Gaussian with variance
$(\sigma/2\Delta)^2$ and $\widetilde{\vw}$ drawn i.i.d. uniform in
$[0,1)$. The distance mapping in~\eqref{eq:ue_dist_map} follows
  from~\eqref{eq:gaussian_dist_map} and the Fourier series of the
  periodic square wave $H_k=\frac{\sin(\pi k/2)}{\pi k}$, exploiting
  the fact that $|\sin(\pi k/2)|^2$ is 0 for $k$ even, and 1 for $k$
  odd. Note, that we are using the Fourier series of a shifted
  quantizer compared to the one in Fig.~\ref{fig:quant}(a)., i.e.,
  $Q'(x)=Q(x+1/2)$, because it is symmetric and has a real Fourier
  series. However, the shift is inconsequential in the result because
  of the dither $\vw$.

Using the above, the distance embedding in~\cite[Thm.~3.2]{BR_WIFS11}
and the kernel approximation in~\cite[Prop.~3.1]{BM_SAMPTA2015} follow
trivially from Thms.~\ref{thm:point_embedding}
and~\ref{thm:kernel_embedding}, respectively. We should also note that
the kernel guarantee is slightly tighter
in~\cite[Prop.~3.1]{BM_SAMPTA2015}, exploiting the fact that a binary
embedding with $\vy$ taking values in $\{-1,1\}$ has deterministic
norm $\|\vy\|_2^2=M$ and not random as is the general case for
Thm.~\ref{thm:kernel_embedding}.

Moreover, in addition to verifying existing results, the development
above provides several generalizations. For example, embeddings using
a matrix with elements drawn from an i.i.d. Cauchy distribution, as
described in~Sec.~\ref{sec:rand-choic}, map the $\ell_1$ distance onto
the hamming space. Again, starting with~\eqref{eq:quantizer_scaling}
and $\mA$ drawn from a Cauchy distribution with scale parameter
$\gamma$, $\widetilde{\mA}$ is Cauchy distributed with scale parameter
$\gamma/2\Delta$. Thus, the embedding map becomes
\begin{align}
  g(d) =
  \frac{1}{2}-\sum_{i=0}^{+\infty}\frac{e^{-\frac{(2i+1)\pi\gamma d}{\Delta}}}
       {\left(\pi(i+1/2)\right)^2},
  \label{eq:l1_dist_map}
\end{align}
where $d$ is the $\ell_1$ distance.

Of course, other mappings are possible by appropriately constructing
the projection matrix \mA, but we omit them here for brevity.

\subsubsection{Extension to Infinite Sets}
\label{sec:extens-infin-sets}
More importantly, using the development in
Sec.~\ref{sec:infinite_discontinuous} and
Thm.~\ref{thm:emb_discontinuous}, it is possible to guarantee binary
universal embeddings on infinite sets, extending the results on finite
point clouds established to-date. In the context of
Thm.~\ref{thm:emb_discontinuous}, the embedding is satisfied with
$c=1$, $\delta=0$ and $w(\epsilon)=2\epsilon^2$. However, the
embedding map $Q(\mA\vx+\vw)$ is discontinuous and, therefore, we need
to examine its $T$-part Lipschitz continuity property.

In particular, since the scalar binary quantizer $Q(\cdot)$ only takes
values in $\{0,1\}$, the embedding function is, at most, exactly
$2$-part Lipschitz continuous with constant $K_f$ equal to $0$. In the
context of Thm.~\ref{thm:emb_discontinuous}, we can bound $P_2\le 1$,
i.e., $c_1\le(1+c_0)\log 2$, and set $P_F=0$. Thus the probability that the
embedding does not hold is upper bounded by
\begin{align}
  ce^{2E_{r/2}^\sS+c_1M-Mw(\delta,\epsilon)}+
  T_{\mathrm{max}}e^{-2c_0^2M}+P_F=e^{2E_{r/2}^\sS+M(1+c_0)\log 2-2M\epsilon^2}+
  2e^{-2c_0^2M}
\end{align}
which decreases exponentially in $M$ as long as
$(1+c_0)\log2<2\epsilon^2$, which only holds if
$\epsilon>\sqrt{0.5\log 2}\approx 0.6$. In other words, this simple
bounding approach can only guarantee an embedding with a large error
$\epsilon$.

A tighter bound can be found if we better understand and bound
$P_2$. This is the probability that a ball of radius $r/2$ will cross
a quantization boundary when projected through a random projection. If
the projected ball diameter is $\Delta$ or greater, then a
quantization boundary will be crossed with probability 1. On the other
hand, if the projected ball diameter is $l\le\Delta$, then a boundary
crossing only happens with probability $l/\Delta$. Thus, a bound on
$P_2$ can be developed which enables the embedding error $\epsilon$ to
go to zero. In the interest of brevity in the core of our development,
we relegate the details on developing the bound to
App.~\ref{sec:prob-cross-univ}.

\subsubsection{Error Analysis}
In contrast to quantized J-L embeddings, binary universal embeddings
use 1 bit per embedding dimension. Thus, the rate $R$ also determines
the dimensionality of the projection, $K=R$, as well as the constant
$\epsilon$ in the embedding
guarantees~\eqref{eq:universal_embedding}. Furthermore, there is no
multiplicative term in the guarantees, i.e., $\delta=0$. Thus, in the
ambiguity analysis~\eqref{eq:ambiguity}, the numerator is fully
determined; the system designer can only control the denominator.

This does not mean that there are no design choices and trade-offs:
the trade-off in these embeddings is in the choice of the parameter
$\Delta$ in~\eqref{eq:universal_quantization}. As discussed in the
Sec.~\ref{sec:univ-quant-embedd} and shown in Fig.~\ref{fig:quant}(b),
$g(\cdot)$ exhibits an approximately linear region, followed by a
rapid flattening and an approximately flat region. The choice of
$\Delta$ controls the slope of the linear region and, therefore, how
soon the function reaches the flat region.

As mentioned earlier, the linear bound
in~\eqref{eq:bound_upper_linear} is a very good approximation of the
upwards sloping linear region of $g(\cdot)$, which has slope
$g'(d)\approx\sqrt{2/\pi}/\Delta$. By decreasing $\Delta$, we can make
that slope arbitrarily high, with a corresponding decrease of the
ambiguity $\epsilon/g'(\widetilde{d}_{\sS})$. However, this linear
region does not extend for all $d$, but only until it reaches the
point $d=D_0$ where $g(D_0)\approx 1/2$ and the flat region of $g(d)$
begins. As $\Delta$ becomes smaller and the slope of the linear region
increases, it reaches the flat region much faster, approximately when
$D_0\sqrt{2/\pi}/\Delta=1/2$, i.e., when
$D_0\approx\Delta\sqrt{\pi/8}\approx 0.6\Delta$.

Unfortunately, beyond that linear region, the slope $g'(d)$ becomes 0
exponentially fast. This implies that the ambiguity
in~\eqref{eq:ambiguity} approaches infinity. Thus, if the embedding
distance $d_{\sW}$ is within $0.5\pm\epsilon$, then it is impossible to
know anything about $d_{\sS}$ by inverting the mapping, other than
$d_\sS\gtrsim D_0$. This makes the trade-off in designing $\Delta$
clear. A smaller $\Delta$ reduces the ambiguity in the range of
distances it preserves, but also reduces the range of distances it
preserves. The system designer should design $\Delta$ such that the
distances required in the application of the embedding are
sufficiently preserved.

As an example, consider the motivating application in
Sec.~\ref{sec:motivation}: retrieval of nearest-neighbors from a
database. When a query is executed, its embedding distance is computed
with respect to all the entries in the database, embedded using the
same parameters. For the query to be successful, there should be at
least a few entries in the database with small embedding distance from
the query. These entries are selected and returned. For the query to
produce meaningful results, the embedding distance of those entries
should represent quite accurately the signal distance between the
query signal and the signals from the entries in the
database. Furthermore, if the signals are all very distant from the
query, the embedding distance should accurately reflect that fact, so
that no signal is selected; in this case the embedding does not need
to represent how distant each entry is.

In other words, the embedding only needs to represent distances up to
a radius $D$, determined by the system designer, and to only {\em
  identify} distances further than $D$, without necessarily
representing those distances. Thus, $\Delta$ should be designed to be
as small as possible so the ambiguity in representing distances in the
linear region is small, but not smaller than necessary to ensure that
all distances of interest are contained in the linear region of the
embedding and do not spill over into the flat region with high
ambiguity.

Note that this notion of locality is much richer that the notion
defined in~\cite{oymak2015near}. The latter only ensures that the
distances between embeddings of close signals is small and between
embeddings of distant signals is large. Instead, our development,
further guarantees a linear distant map up to a radius, thus
preserving distances up to this radius.

%% file: ex_multibituniversal.tex
\subsection{Multibit Universal Embeddings}
\label{sec:uq_multibit_theory}
Another benefit of the development and analysis in this paper is that
it facilitates analysis of multibit universal embeddings, i.e.,
embeddings using the quantizer at the bottom of
Fig.~\ref{fig:quant}(a).

In principle, it is possible to consider multi-bit universal
quantizers as sums of scaled one-bit quantizers, both in amplitude and
the argument. In particular, given the 1-bit quantizer $Q(\cdot)$
defined at the top of Fig.~\ref{fig:quant}(a), the $B$-bit
generalization equals
\begin{align}
  Q_B(x)=\sum_{b=0}^{B-1} 2^b Q\left(\frac{x}{2^b}\right),
\end{align}
and has period $2^B$. Thus, with quantization interval $\Delta$, the
embedding can be expressed as
\begin{align}
  \vy=\widetilde{Q}_B\left(\widetilde{\mA}\vx+\widetilde{w}\right),
\end{align}
where $\widetilde{Q}_B(x)=Q_B(2^Bx)$, $\widetilde{\mA}$ has elements
drawn from an i.i.d. Normal distribution with variance
$(\sigma/2^B\Delta)^2$, to map $\ell_2$ distances, or an i.i.d. Cauchy
distribution with scale parameter $\gamma/2^B\Delta$ to map $\ell_1$
distances, as described above. Using the Fourier series of $Q(\cdot)$,
appropriately scaled and summed for each bit $b=0,\ldots,B-1$, and a
similar development as in Sec.~\ref{sec:embedding-map} it is possible
to derive the embedding map. It is important to note, however, that
appropriate shifting and scaling of the functions significantly
complicates the resulting expressions.

Instead, a simpler expression can be derived by observing that the
multibit universal quantization function is, in fact, the result of
uniform scalar quantization applied to a sawtooth map. Thus, we can
use the well-established Fourier series coefficients of the sawtooth
function to obtain $|H_k|^2=(1/\pi k)^2$ for $k>0$. The resulting map
for a general \mA\ without scaling and quantization equals to
\begin{align}
  g(d)=2\sum_{k>0}\frac{1}{\pi^2 k^2}(1-\phi_l(2\pi
  k|d))=\frac{1}{3}-2\sum_{k>0}\frac{1}{\pi^2 k^2}\phi_l(2\pi k|d).
  \label{eq:sawtooth_map}
\end{align}
Thus, if the scaling by $\Delta$ and $2^B$ is considered and the
elements of \mA\ are drawn from an i.i.d. $\mathcal{N}(0,\sigma^2)$
distribution, the embedding map becomes
\begin{align}
  g(d)=\frac{1}{3}-2\sum_{k>0}\frac{1}{\pi^2
    k^2}e^{-2\left(\frac{\pi\sigma dk}{2^B\Delta}\right)^2},
\end{align}
where $d$ is the $\ell_2$ distance of the signals. Similarly, if
elements of \mA\ are drawn from an i.i.d. Cauchy distribution with
scale parameter $\gamma$, then the embedding map becomes
\begin{align}
  g(d)=\frac{1}{3}-2\sum_{k>0}\frac{1}{\pi^2 k^2}e^{-\frac{2\pi\gamma
      dk}{2^B\Delta}},
\end{align}
where $d$ is now the $\ell_1$ distance of the signals.

The sawtooth, which takes values in $[-1/2,1/2]$ is subsequently
quantized by a $B$-bit uniform scalar quantizer, i.e., one having
$2^B$ levels and interval $\Delta=2^{-B}$. In the context of
Thm.~\ref{thm:quantized_embedding}, this implies a worst-case
quantization error of $E_Q=\Delta/2=2^{-B-1}$. This error is in the
$\ell_2$ distance of the embedding, not in the $\ell_2^2$ guarantee of
Thm.~\ref{thm:point_embedding}. The guarantee applies, therefore, to
the embedding map in Cor.~\ref{cor:root_point_embedding} and
Cor.~\ref{cor:root_point_embedding_tight}. The corresponding guarantee
for the $\ell_2^2$ distance should use $E_Q=(\Delta/2)^2=2^{-2B-2}$.

Using a sawtooth function as a map, followed by scalar quantization to
derive multibit universal embedding guarantees should provide looser
bounds than explicitly estimating the guarantee using sum of square
wave maps. However, as the rate $B$ increases, the two guarantees
should converge to the same one. In fact as $B$ increases and
quantization becomes finer, both approaches converge to the guarantees
derived using an unquantized sawtooth function as an embedding map.

%% file: simulations.tex
\section{Simulations and Application Examples}
\label{sec:simul-appl-exampl}
To verify and demonstrate the theoretical developments above, we
present simulation results verifying our designs. We also demonstrate
how our approach can be applied toward encoding features for image
retrieval and for image classification over the network.

\input{sa_design}
\input{sa_nn_retrieval}

\input{sa_kernel_classification}

%% file: sa_design.tex
\subsection{Simulations: Embedding Design}
Existing simulation results on quantized embeddings, such as the ones
shown in Fig.~\ref{fig:quant}, demonstrate the validity of our
analysis. To further verify the results in Sec.~\ref{sec:design} we
consider a slightly more complex distance map, in which shorter
distances should be represented with greater precision, intermediate
distances should be represented with less precision and larger
distances do not need to be represented with any precision, similar to
universal embeddings.

\begin{figure}[t]
  \centerline{\includegraphics[width=.9\linewidth]{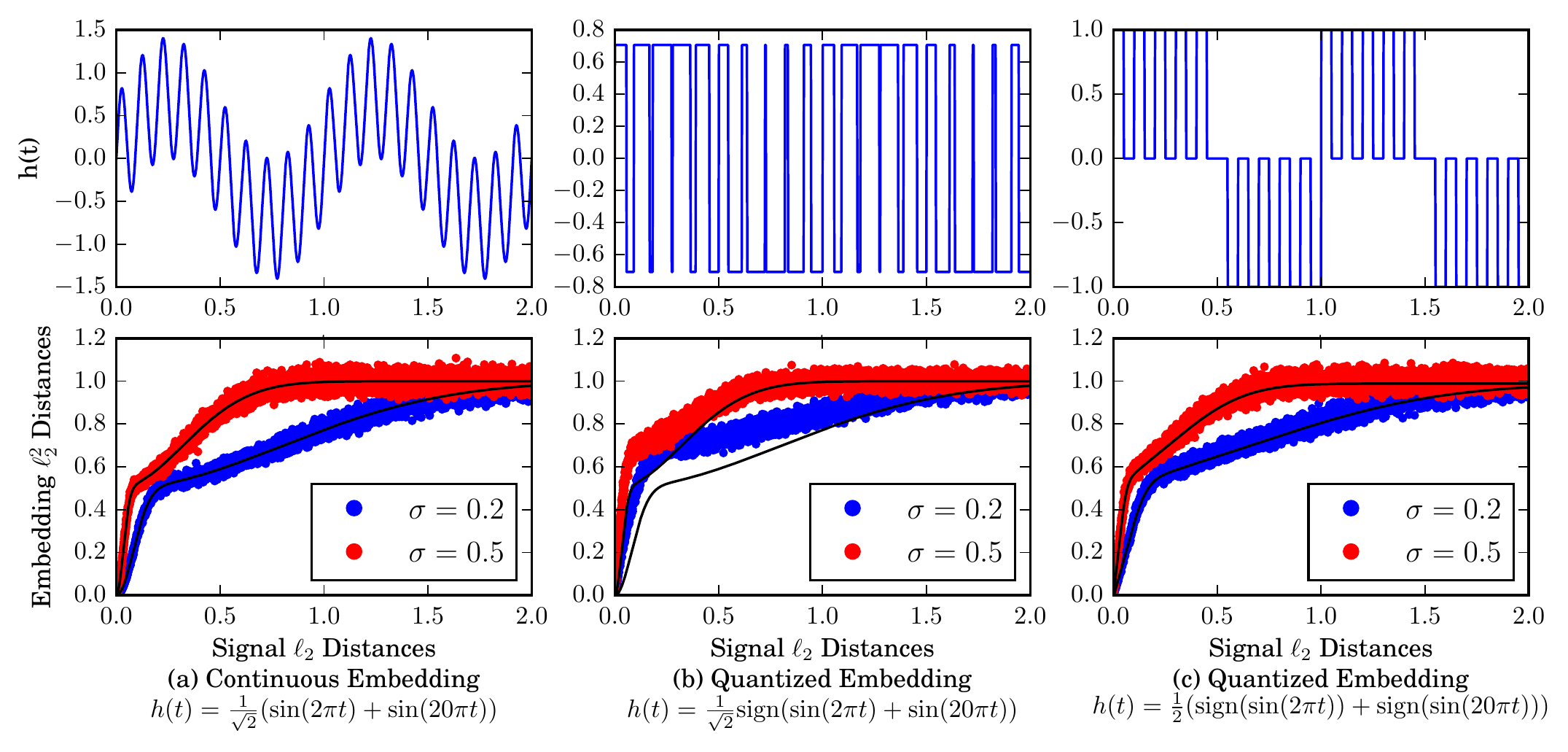}}
  \caption{Simulation results in embedding design: (a) Unquantized
    embedding preserving two different distance intervals with
    different accuracy. (b) Effect of 1-bit quantization on (a). (c) A
  3-level quantized embedding with similar performance as (a).}
  \label{fig:design_simulation}
\end{figure}

Specifically, we first consider an embedding with
$H_1=H_{10}=\sqrt{2}/2$ and $H_k=0$ for all other $k$. In other words,
the embedding map $h(t)$ is equal to
\begin{align}
  h(t)=\frac{\sqrt{2}}{2}(\sin(2\pi t)+\sin(20\pi t)),
  \label{eq:embedding_example}
\end{align}
as plotted at the top of Fig.~\ref{fig:design_simulation}(a). 

The bottom of Figure~\ref{fig:design_simulation}(a) demonstrates the
performance of the embedding on randomly generated signals in
$N=10000$ dimensions with different distances in the range
$d=0,\ldots,2$. The signals are embedded in $M=2000$ dimensions using
matrices with variance $\sigma=0.2$ and $0.4$ for the blue and red
dots, respectively. The $\ell_2^2$ distance of the embedded signals is
plotted against the $\ell_2$ distance of the signals. The black line
in the figure plots the theoretical embedding map $g(d)$ according to
Thm.~\ref{thm:point_embedding}. As evident in the figure, the
embedding performs as predicted by the theory. The embedding map
increases rapidly for a small radius, not as rapidly until a larger
radius, and then becomes flat beyond that. The radii are greater for
smaller $\sigma$, as expected.

The figure also shows the ambiguity, as measured by the horizontal
width of the plots and analyzed in Sec.~\ref{sec:error-analysis}. As
expected, the ambiguity is lower given the vertical ambiguity if the
slope of the embedding map is higher. Thus, short distances, up to a
first radius, are better represented than longer distances, up to the
point where the embedding flattens. Beyond that, the embedding only
conveys information that signals are far apart.

We should also note, that it seems that the vertical ambiguity of the
embedding seems to be smaller for smaller embedding distances,
suggesting a multiplicative ambiguity instead of an additive one. 

Figure~\ref{fig:design_simulation}(b) demonstrates a 1-bit quantized
version of the embedding in (a), simply taking the sign of the
continuous embedding:
\begin{align}
  h(t)=\frac{\sqrt{2}}{2}\sign(\sin(2\pi t)+\sin(20\pi t)),
\end{align}
with the scaling chosen such that $g(d)$ saturates to 1
asymptotically. The embedding map is shown on the top of the figure,
while the embedding performance is shown at the bottom. The black line plots
the theoretical embedding map for the unquantized embedding, i.e., the
same map as in (a). 

As evident by the figure, the actual performance of the embedding
concentrates around a curve that is different than the theoretical
prediction for the continuous version. However, the embedding is
within the bounds of Thm.~\ref{thm:quantized_embedding} because $E_Q$
is quite large for a 1-bit quantizer. Moreover, the experimental
results demonstrate quite good concentration around a curve, even
though the embedding uses only one bit per coefficient. Still, a
better understanding of the quantized embedding map and its Fourier
series would yield a more accurate prediction. Such an understanding
is not straightforward and we do not attempt it here.

Instead, in Figure~\ref{fig:design_simulation}(c) we demonstrate a
similar quantized embedding that is easier to
characterize. Specifically, the embedding map $h(t)$ is equal to
\begin{align}
  h(t)=\frac{1}{2}\sign(\sin(2\pi t))+\frac{1}{2}\sign(\sin(20\pi t)),
\end{align}
where
\begin{align}
  \sign(x)=\left\{
  \begin{array}{rl}
    -1,&\mathrm{if}~x<0\\
    +1,&\mathrm{otherwise,}
  \end{array}
  \right.
\end{align}
as plotted at the top of the figure. In this case,
\begin{align}
  |H_k|=\left\{
  \begin{array}{rl}
    2/\pi k,&\mathrm{if}~k~\mathrm{is~odd}\\
    20/\pi k,&\mathrm{if}~k~\mathrm{is~divisible~by}~10\\
    0,&\mathrm{otherwise.}
  \end{array}
  \right.
\end{align}
This embedding, as shown in the figure, has very similar
characteristics and distance-preservation properties with the
continuous embeddings of Fig.~\ref{fig:design_simulation}(a). The
quantized embedding has slightly wider error bounds in the
experiments, but this is expected given that it is quantized at only 3
levels per coefficient. Still, despite the effect of quantization, the
performance is very close to the performance of the continuous
embedding. In contrast to the quantized embedding in
Fig.~\ref{fig:design_simulation}(b), the embedding in (c) can be
easily analyzed. As expected, it is also a bit tighter than the one in
(b) because it is a 3-level per coefficient embedding, i.e., uses
approximately $1.6\times$ the rate.

\begin{figure}[t]
  \centerline{\includegraphics[width=.75\linewidth]{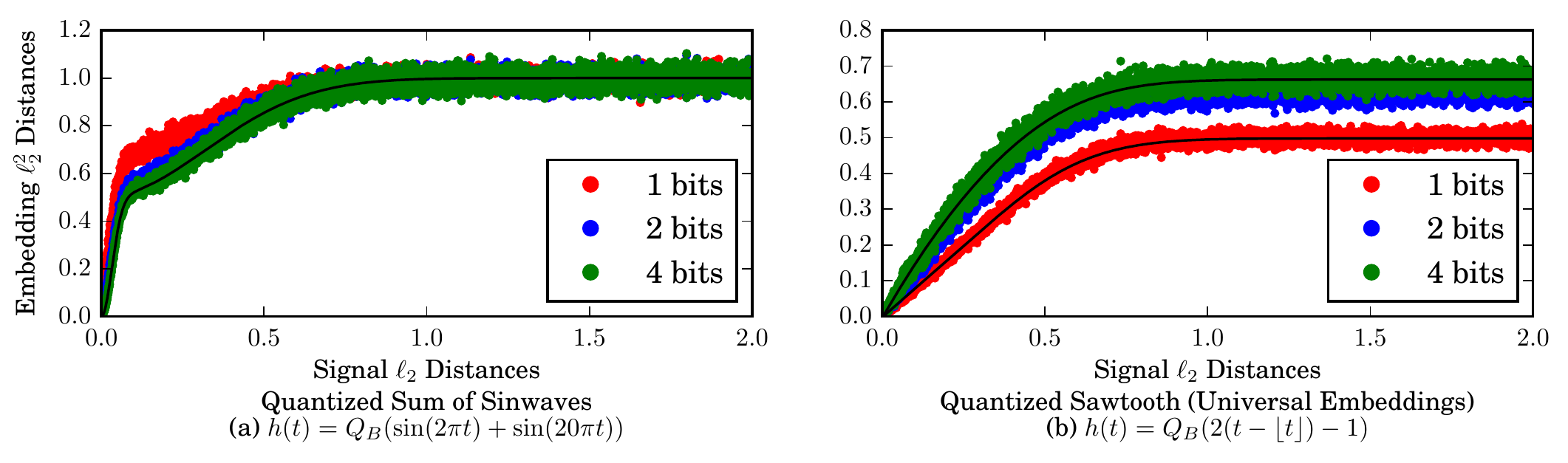}}
  \caption{Effect of quantization on embeddings. (a) The embedding of
    Fig.~\ref{fig:design_simulation}(a) with increasingly refined
    quantization. (b) Multibit universal embeddings with increasingly
    refined quantization, approaching the performance of a continuous
    embedding using a sawtooth embedding map.}
  \label{fig:quantization_simulation}
\end{figure}

To further demonstrate the effect of quantization,
Fig.~\ref{fig:quantization_simulation}(a) plots simulation results on the
embedding map~\eqref{eq:embedding_example} when quantized with a 1-,
2-, and 4-bit scalar quantizer, i.e., for embedding maps of the form
\begin{align}
  h(t)=Q_B(\sin(2\pi t)+\sin(20\pi t)),
\end{align}
where $Q_B(\cdot)$ is a $B$-bit scalar quantizer, appropriately
scaled. As above, the black line plots the distance map predicted
assuming an unquantized embedding, i.e., the embedding in
Fig.~\ref{fig:design_simulation}(a). Of course, the case $B=1$
coincides with the example in Fig.~\ref{fig:design_simulation}(b).

As evident in Fig.~\ref{fig:quantization_simulation}(a), the higher
the rate, the closer the result is to the theoretical prediction for the
continuous embedding. Of course, the embedding satisfies the bounds of
Thm.~\ref{thm:quantized_embedding}. However, the figure suggests that
the bounds are loose. An analysis along the lines of
Thm.~\ref{thm:point_embedding}, should provide a tighter bound. Such
an analysis is not as straightforward for this particular embedding,
and we do not attempt it here.

For completeness, Fig.~\ref{fig:quantization_simulation}(b) plots the
performance of quantized universal embeddings for the same choice of
bitrates. The figure also plots the theoretical distance
maps~\eqref{eq:universal_embedding} for $B=1$
and~\eqref{eq:sawtooth_map} for the unquantized sawtooth wave using
black lines. Even with $B=4$, the quantization is sufficiently fine, so that
the experimental distance map of the quantized embedding and the
theoretical distance map of unquantized embedding are observed to coincide.

%% file: sa_nn_retrieval.tex
\subsection{Application Example: Image Retrieval Using Universal Embeddings}
\label{sec:appl-exampl-image}

As an example application we consider image retrieval over the cloud.
A user wants to retrieve information about a query object by capturing
its photograph and transmitting information extracted from the
photograph to a database server. The server locates the object in the
database that most closely matches the query image, according to a
predetermined similarity criterion, and transmits meta-data about that
object back to the user. The goal is to reduce, as much as possible,
transmission bit-rate given a certain desired performance. 

Since the server does not require to exactly reconstruct the image to
retrieve similar images, it should be possible to significantly reduce
the bit-rate compared to naively transmitting the actual images using
lossy compression. As we describe below, this is indeed possible by
computing quantized universal embeddings of features extracted from
the query and database images.

\subsubsection{Protocol Architecture}
\label{sec:prot-arch-nn}
In preparation for the query, server and  client agree on the
embedding parameters---specifically, \mA, \vw, and $\Delta$ in the
case of universal embeddings---according to the embedding
specifications. For example, the server might draw universal
parameters for all clients that will access the database.

Next, the server builds a database using features extracted from
previously labeled images. In our experiments we use the
well-established SIFT features~\cite{SIFT}, which provide significant
invariance properties that facilitate image retrieval. Typically, with
such feature extraction methods, a single image might generate a
variable number of features. However, each of the generated feature
vectors is associated to an image in the database and its associated
metadata. The server builds and indexes the database by embedding the
features using the predetermined embedding parameters and associating
the image and the metadata with the correct embedded feature.

To execute a query, the client first acquires an image that serves as
the query image. The client extracts the features from that image,
embeds them using the predetermined embedding parameters, and
transmits their embedding to the server. The server receives the
embedded features, and retrieves from the database the nearest
neighbor to each feature using the embedding distance, i.e., a single
match for every embedded features. From those matches, the server
selects the $J$ closest candidates, with $J=20$ in our experiments. The metadata are
selected using majority voting among those matches.

\input{results}

%% file: results.tex
\subsubsection{Experimental Results}
\label{sec:expts}
\begin{figure*}[t]
\begin{minipage}[l]{.3\linewidth}
  \begin{center}
    \includegraphics[width=.895\linewidth]{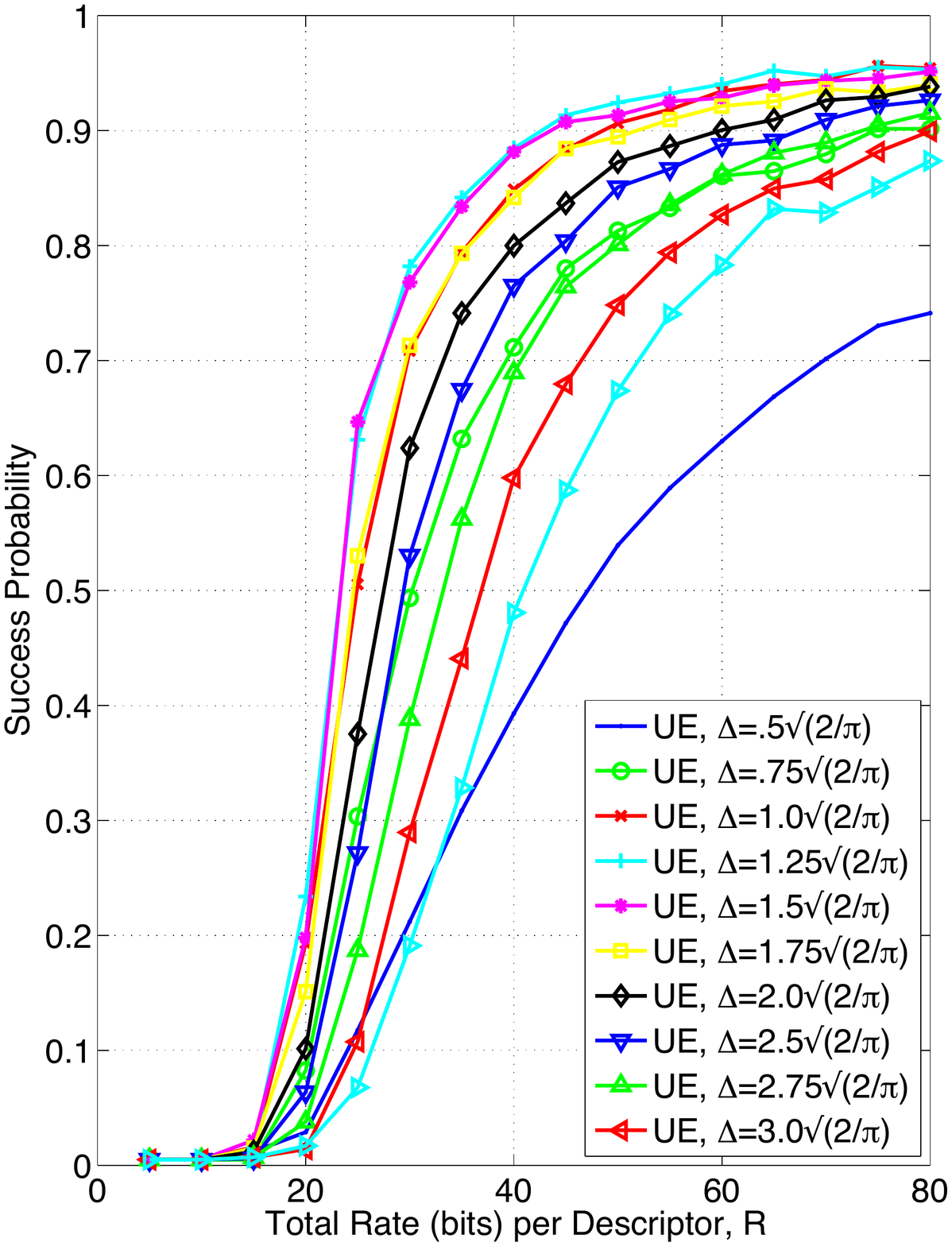}\\
  \end{center}
\end{minipage}
\hfill
\begin{minipage}[c]{.3\linewidth}
  \begin{center}
    \includegraphics[width=.895\linewidth]{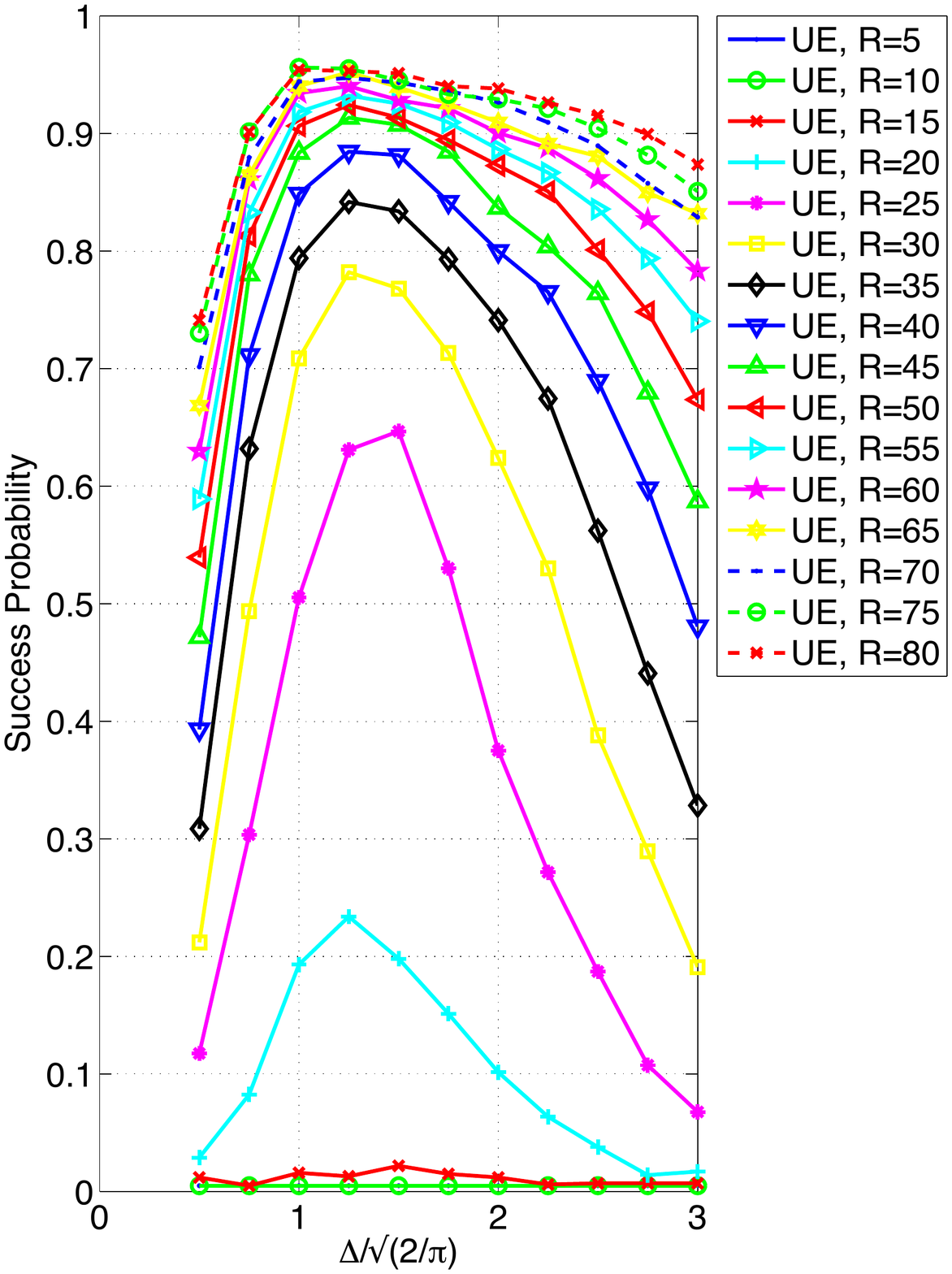}\\
  \end{center}
\end{minipage}
\hfill
\begin{minipage}[r]{.3\linewidth}
  \begin{center}
    \includegraphics[width=.895\linewidth]{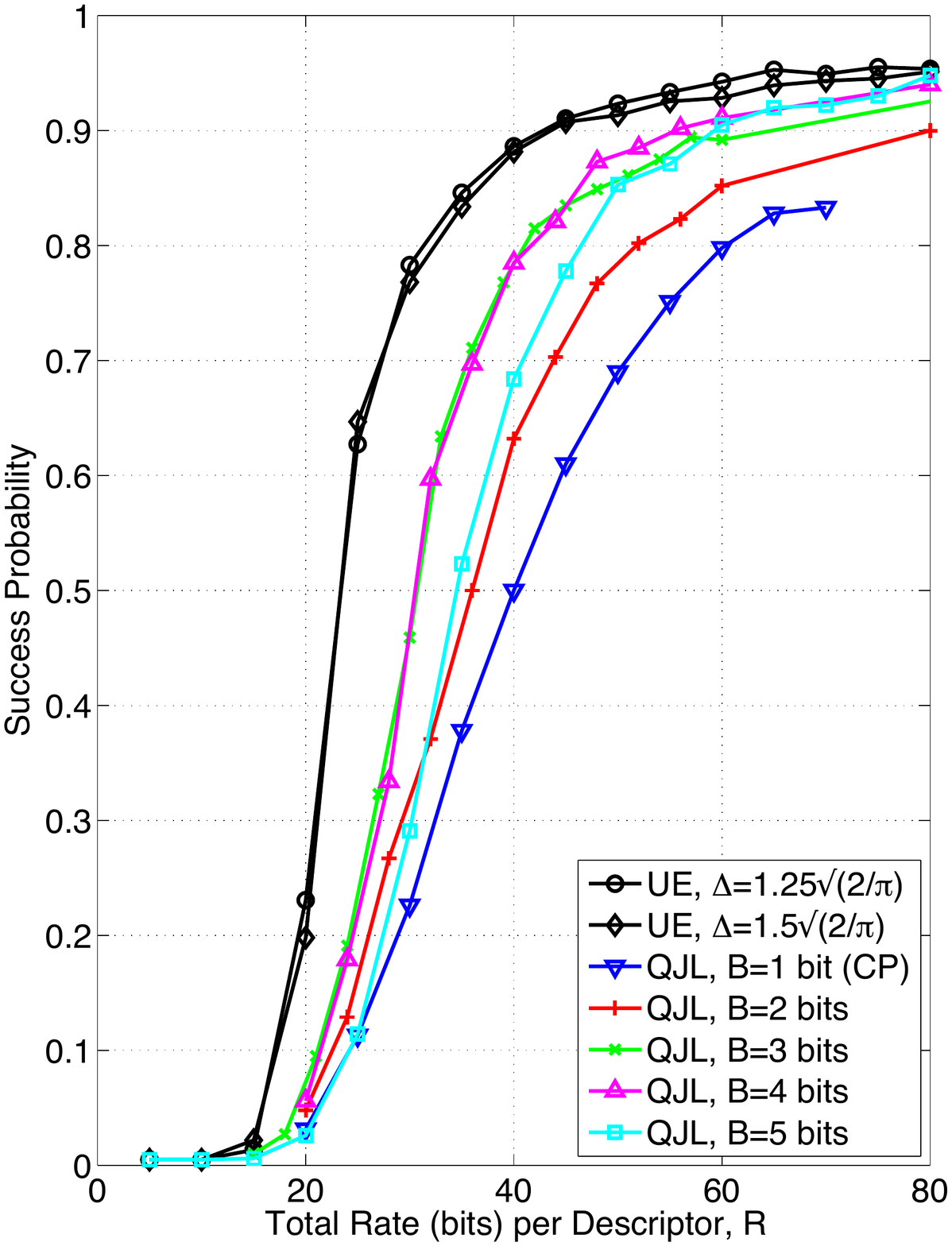}\\
  \end{center}
\end{minipage}
\\
  \begin{minipage}[c]{.3\linewidth}
    \begin{center}
      \footnotesize{(a)}
    \end{center}
  \end{minipage}\hfill
  \begin{minipage}[c]{.3\linewidth}
    \begin{center}
      \footnotesize{(b)}
    \end{center}
  \end{minipage}\hfill
  \begin{minipage}[c]{.3\linewidth}
    \begin{center}
      \footnotesize{(c)}
    \end{center}
  \end{minipage}
\caption{Performance of universal embeddings (UE) in metadata
  retrieval. (a) Probability of correct retrieval as a function of the
  bitrate for a variety of $\Delta$ values. (b) Probability of correct
  retrieval as a function of $\Delta$ for a variety of bitrates. (c)
  Comparison of universal embeddings using $\Delta=1.25\sqrt{2/\pi}$
  and $1.5\sqrt{2.\pi}$ with quantized J-L methods (QJL). Universal
  embeddings significantly outperform the alternatives.}
\label{fig:results}
\end{figure*}

To validate our approach, we conducted retrieval experiments using the
ZuBuD database~\cite{ZuBuD}. This public database contains 1005 images
of 201 buildings in the city of Zurich. There are 5 images of each
building taken from different viewpoints, all of size $640
\times 480$ pixels and compressed in PNG format.  Our experimental
setup is identical to~\cite{LRB_MMSP12}: One out of the 5 viewpoints
of each building was randomly selected as a query image, forming a
test set of $s=201$ images. The server's database comprises of the
remaining 4 images of each building, for a total of $t=804$
images. The query aims to identify which of the 201 possible buildings
is depicted in each query image.

Our goal is to examine the performance of embeddings in preserving
distances, not the performance of various feature selection methods or
retrieval protocols. Thus, we extracted the widely adopted SIFT
features~\cite{SIFT} from each image and embedded them using quantized
J-L embeddings or universal embeddings. Using the protocols described
in Sec.~\ref{sec:prot-arch-nn} we measured how many of the 201 query
images produced the correct result, i.e., correctly identified the
building depicted. We conducted our experiments in bitrates ranging
from 0 to 80 bits per descriptor. Our results are averaged over 100
experiments with different realizations of \mA\ and \vw, although the
variability among individual runs was very small.

The first experiment tested the effect of $\Delta$ in the design of
the embedding. In particular, we examined the range
$\Delta=0.5\sqrt{2/\pi},0.75\sqrt{2/\pi},\ldots,3\sqrt{2/\pi}$. The
results are shown in Figs.~\ref{fig:results}(a) and (b). In
Fig.~\ref{fig:results}(a) each curve plots the probability of correct
metadata retrieval as a function of the bitrate used per descriptor,
given a fixed $\Delta$. The higher the probability of success, the
better. Figure~\ref{fig:results}(b) presents another view on the same
data: each curve plots the probability of correct retrieval given a
fixed bitrate per descriptor as $\Delta$ varies.

The plots in Fig.~\ref{fig:results}(a) and (b) verify our
expectations. As the bitrate increases, the performance improves. With
respect to $\Delta$, the behavior is more nuanced. For small $\Delta$,
the slope of $g(d)$ is high and the ambiguity in the linear region of
$g(d)$ is low, as discussed in Sec.~\ref{sec:uq_theory} and shown in
Fig.~\ref{fig:quant}. Thus, the distances represented by the embedding
are represented very well. However, $D_0$ is small, i.e. it can only
represent accurately a very small range of distances. Thus, for a
large number of queries for which the closest matches are farther than
$D_0$ the results returned are not meaningful. This type of error
dominates the results when $\Delta$ is low. As $\Delta$ increases,
more and more queries produce meaningful results and the error
performance improves, even though the accuracy of the linear region of
the embedding decreases. For larger $\Delta$ the reduced accuracy of
the embedding starts dominating the error and the performance
decreases again. The best performance is obtained for
$\Delta=1.25\sqrt{2/\pi}$, which corresponds to corresponding
$D_0=.625$.

We also compared the performance of our approach using quantized J-L
embeddings. Figure~\ref{fig:results}(c) compares the performance of
the two types of embeddings. The figure plots the probability of
correct retrieval as a function of the bitrate per descriptor for each
of the methods examined. As expected~\cite{LRB_MMSP12}, multibit
quantized J-L embeddings outperform 1-bit quantized J-L
embeddings---known as ``compact projections'' (CP)~\cite{RP,CP} and
motivated by LSH approaches~\cite{LSH} in earlier literature. More
important, universal embeddings---plotted in black circles and black
diamonds, for $\Delta=1.25\sqrt{2/\pi}$ and $1.5\sqrt{2/\pi}$
respectively---significantly outperform quantized J-L embeddings. For
example, to achieve a probability of correct retrieval of $80\%$,
universal embeddings require approximately 8 fewer bits per
descriptor, a $20\%$ rate reduction. For $90\%$ probability of correct
retrieval, universal embeddings require 15 fewer bits per descriptor,
a $25\%$ rate reduction. Similarly, using only 40 bits per descriptor,
universal embeddings achieve almost $90\%$ success rate, versus almost
$80\%$ for the best alternative. The results are robust with respect to $\Delta$:
for $\Delta\in[\sqrt{2/\pi},2\sqrt{2/\pi}]$, universal embeddings
outperform all quantized J-L embeddings.

%% file: sa_kernel_classification.tex
\subsection{Application Example: Kernel-based Image classification}
\label{sec:appl-kern-based}
We also tested the performance of universal embeddings as a feature
representation on a multiclass classification problem. The goal is to
identify the class membership of query images belonging to one of 8
different classes. Our experiments demonstrate how the embeddings
function as an SVM kernel, and enable a rate-inference trade-off
analogous to the rate-distortion trade-off in conventional coding:
they allow trading inference performance for reduction in the rate.

\subsubsection{Protocol Architecture}
\label{sec:prot-arch-svm}
The protocol for this application is very similar to the one in
Sec.~\ref{sec:prot-arch-nn}. Specifically, to set up this problem, the
server extracts a Dalal-Triggs Histogram of Oriented Gradients (HOG)
features~\cite{DT05} from the training images. The HOG algorithm
extracts a 36 element feature vector (descriptor) for every $8\times
8$ pixel block in an image. The descriptors encode local 1-D
histograms of gradient directions in small spatial regions in an
image. Every HOG feature is compressed using either quantized JL
embeddings or universal quantized embeddings. The compressed features
are then stacked to produce a single compressed feature vector for
each image. The compressed features of the training images are used
together with the image labels to train a number of binary linear SVM
classifiers, one for each class.

To classify a query image, the client extracts HOG features, encodes
them using the predetermined embedding parameters, and transmits it to
the server. The server executes the SVMs directly on the embeddings
and decides on the embedding class.

\subsubsection{Experimental Results}
In our simulations, we used tools from the VLFeat
library~\cite{vedaldi08vlfeat} to extract HOG features and train the
SVM classifier. We consider eight image classes. One class consists of persons
from the INRIA person dataset~\cite{DT05,INRIAdataset}. The other
seven classes---car, wheelchair, stop sign, ball, tree, motorcycle, and
face--- are extracted from the Caltech 101
dataset~\cite{FeiFei04,Caltech101}. All images are standardized to
$128\times 128$ pixels centered around the target object in each
class. We use 15 training and 15 test images from each class.

\begin{figure}
 \begin{minipage}[c]{.3\linewidth}
\centering
\includegraphics[width = \textwidth]{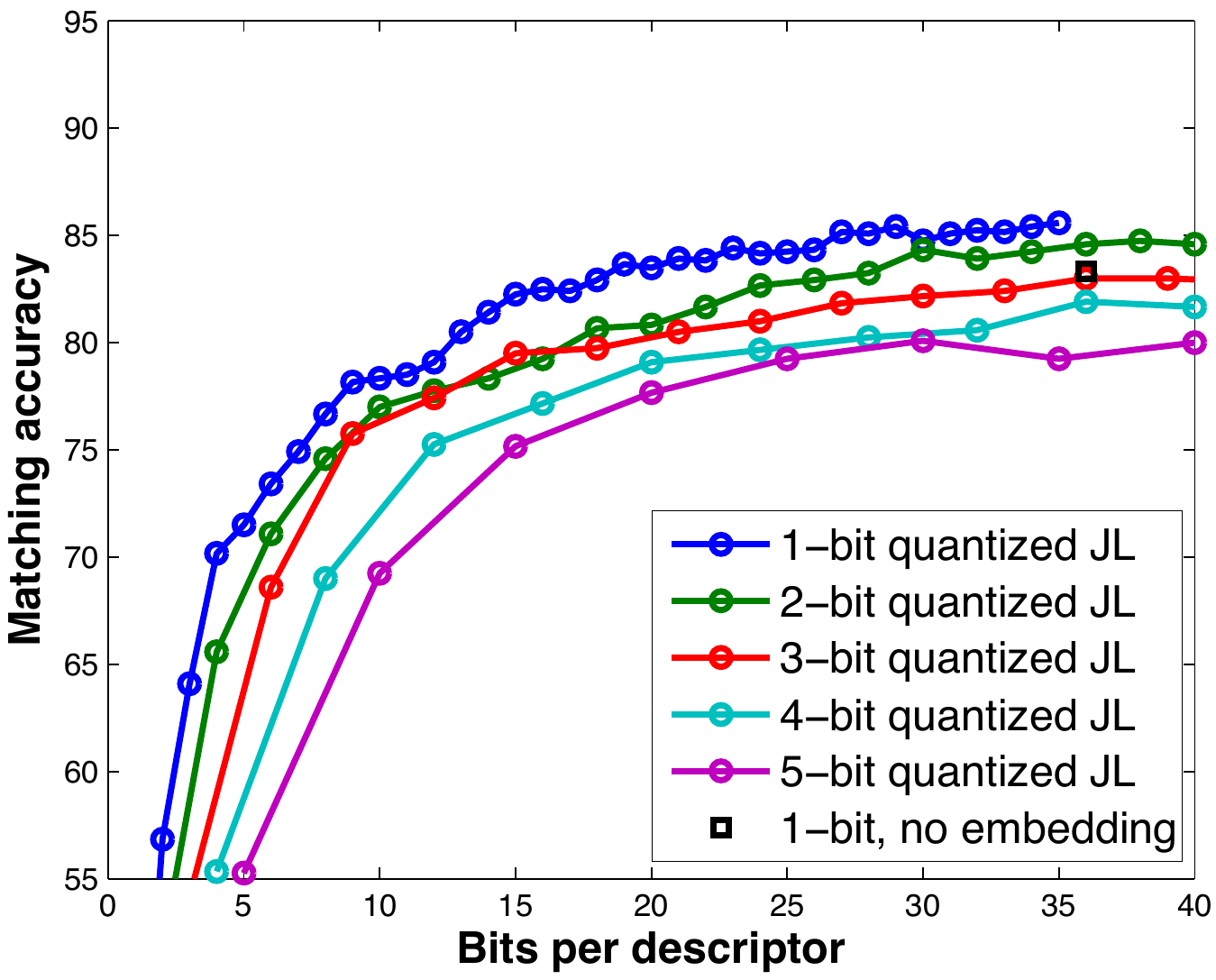}
\end{minipage}
\hfill\begin{minipage}[c]{.3\linewidth}
\centering
\includegraphics[width = \textwidth]{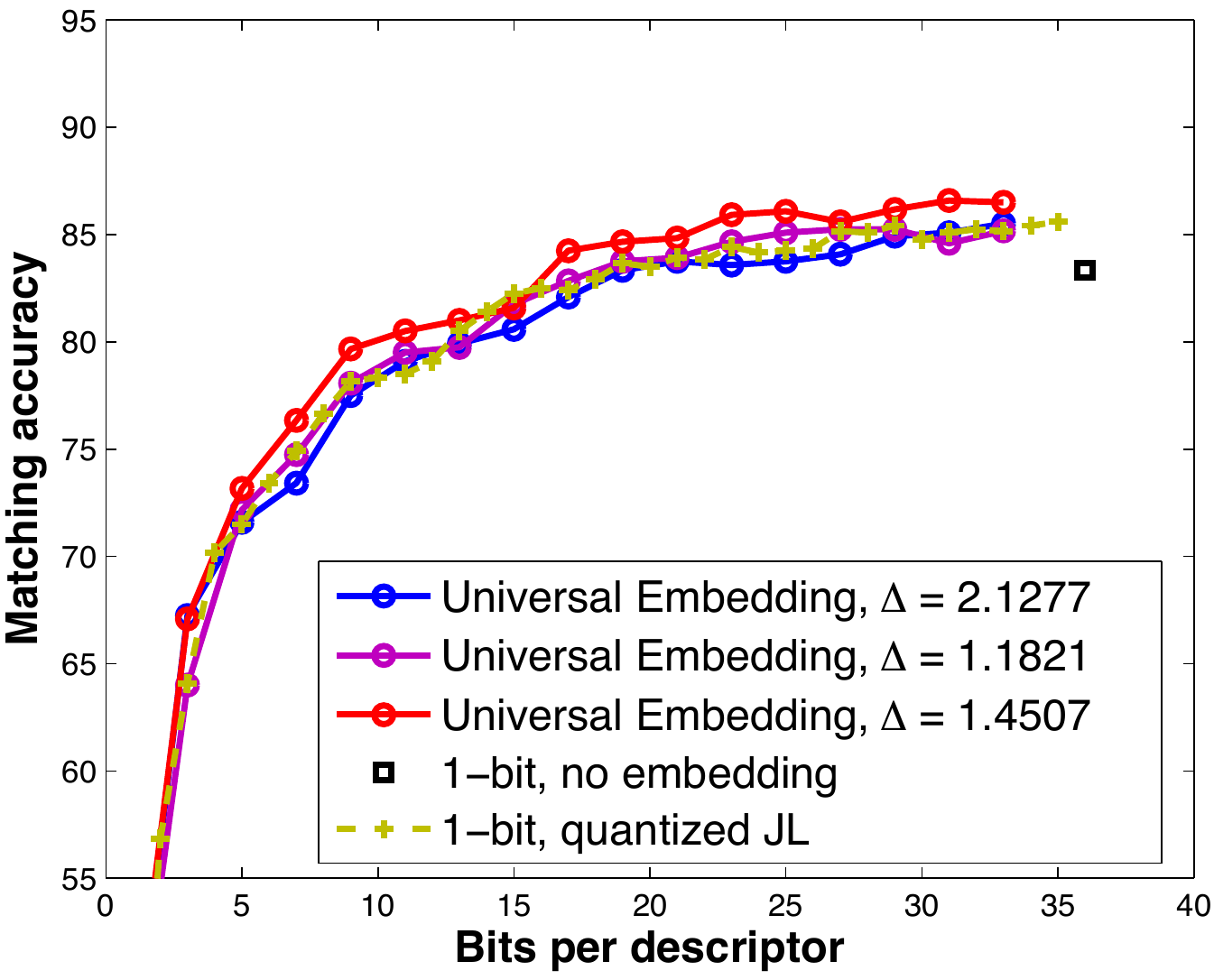}
\end{minipage}
\hfill\begin{minipage}[c]{.3\linewidth}
\centering
\includegraphics[width = \textwidth]{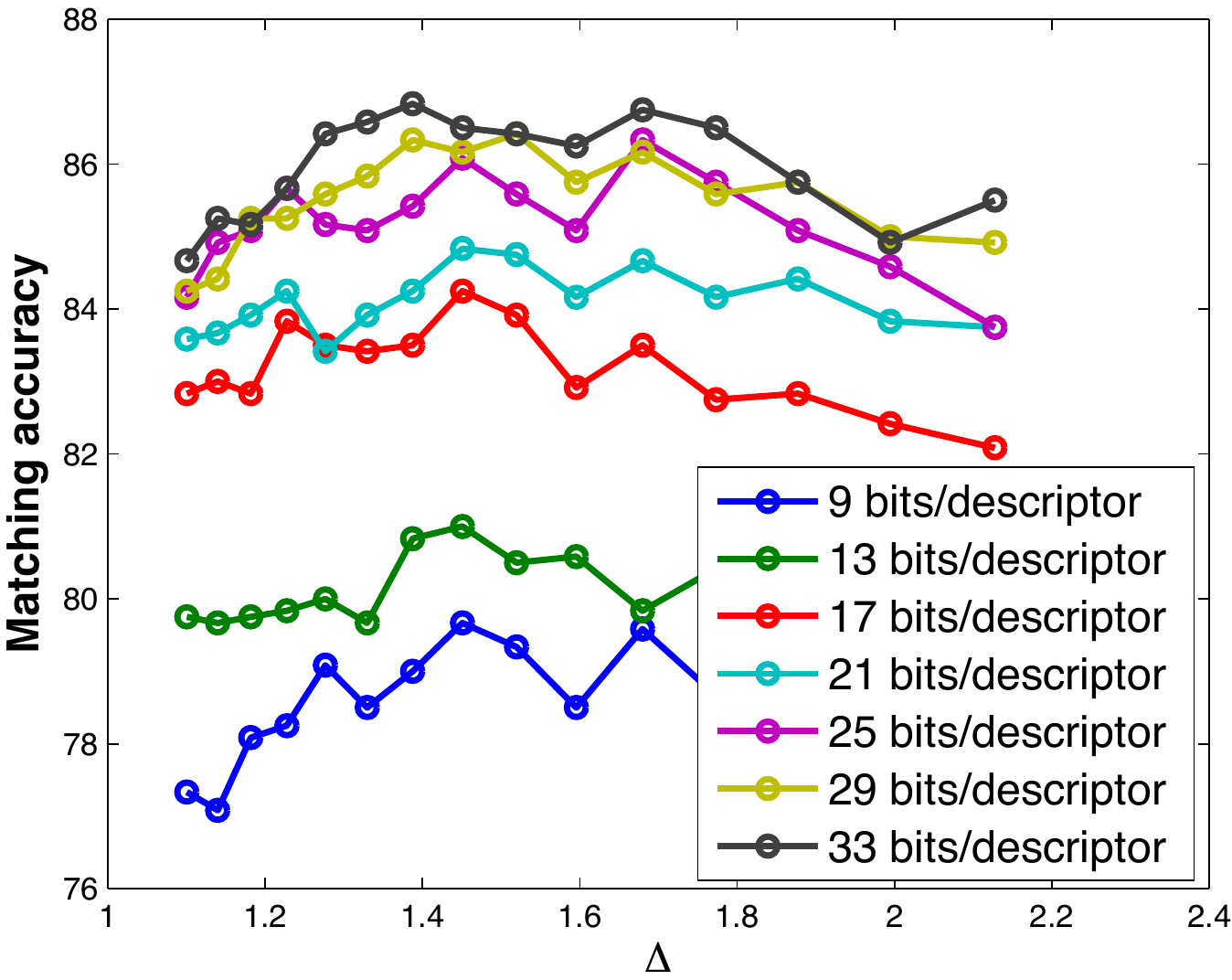}
\end{minipage}
\\
  \begin{minipage}[c]{.3\linewidth}
    \begin{center}
      \footnotesize{(a)}
    \end{center}
  \end{minipage}\hfill
  \begin{minipage}[c]{.3\linewidth}
    \begin{center}
      \footnotesize{(b)}
    \end{center}
  \end{minipage}\hfill
  \begin{minipage}[c]{.3\linewidth}
    \begin{center}
      \footnotesize{(c)}
    \end{center}
  \end{minipage}
\caption{Classification accuracy as a function of the bit-rate
  achieved using (a) quantized JL (QJL) embeddings; and (b) universal
  embeddings. (c) Classification accuracy as a function of the
  quantization step size $\Delta$ used in computing the universal
  embeddings.}\label{fig:Classification}
\end{figure}

Fig.~\ref{fig:Classification}(a) shows the classification accuracy
obtained by quantized JL embeddings of HOG descriptors using the
trained SVM classifier. The black square corresponds to 1-bit scalar
quantization of raw non-embedded HOG descriptors, using a bit-rate of
36 bits---one bit for each element of the descriptor.

The figure shows that 1-bit quantized JL embeddings allow us to
achieve a 50\% bit-rate reduction, compared to non-embedded quantized
descriptors, without reduction in performance (classification
accuracy). This is obtained using an 18-dimensional JL embedding of
every HOG descriptor, followed by 1-bit scalar
quantization. Furthermore, increasing the embedding dimension, and,
therefore, the bit-rate, above 18 improves the inference performance
beyond that of the 1-bit quantized non-embedded HOG features. Note
that, among all quantized JL embeddings, 1-bit quantization achieves
the best rate-inference performance.

Fig.~\ref{fig:Classification}(b) compares the classification accuracy
of universal embeddings for varying values of the step size parameter
$\Delta$ with that of the 1-bit quantized JL embeddings and the 1-bit
quantized non-embedded HOG descriptors. With the choice of $\Delta =
1.4507$, the universal embedded descriptors further improve the
rate-inference performance over the quantized JL embeddings by $~1\%$
in inference improvement. They also achieve the same classification
accuracy as any choice of quantization for non-embedded HOG
descriptors, or, even, unquantized ones, at significantly lower
bit-rate---points not shown in the figure, as they are out of the
interesting part of the bit-rate scale.

Figure~\ref{fig:Classification}(c) illustrates the effect of the
parameter $\Delta$ by plotting the classification accuracy as a
function of $\Delta$ for different embedding rates. The figure shows
that, similar to the findings in Sec.~\ref{sec:appl-exampl-image}, if
$\Delta$ is too small or too large, the performance suffers.

As evident, an embedding-based system design can be tuned to operate
at any point on the rate vs. classification performance frontier, not
possible just by quantizing the raw HOG features. Furthermore, with
the appropriate choice of $\Delta$, universal embeddings improve the
classification accuracy given the fixed bit-rate, compared with
quantized JL embeddings, or reduce the bit-rate required to deliver a
certain inference performance.

%% file: discussion.tex
\section{Discussion and Conclusions}
\label{sec:discussion}
A key contribution of our paper is the notion that embeddings can be
designed to have different distance preservation accuracy for
different distance ranges. In doing so, we developed a framework to
understand how the properties of the embeddings are preserved,
exploiting the embedding distance map. However, our work raises more
interesting questions than it solves.

A key question in our approach is whether any arbitrary distance
preservation design is possible. Our results on the subadditivity of
the distance map in Sec.~\ref{prop:subadditivity} place some
constraints on what distance maps can be realized. However, it is not
clear that this is the only constraint that is realizable. For
example, we conjecture that the distance map should be
monotonic---within $\epsilon$ and $\delta$ ambiguities, similar to the
subadditivity constraint---but we have not provided such a
proof.

It is also not clear that our design method in Sec.~\ref{sec:design}
can achieve any arbitrary design within those constraints. For
example, our design is monotonic. If our monotonicity conjecture is
false, then our design approach cannot achieve all the possible
distance maps. Ideally, we prefer to be able to determine the
embedding starting from the distance map instead of the other way
around that Thm.~\ref{thm:point_embedding} establishes. In the context
of our design approach, we would like to determine $h(t)$ and the
corresponding $H_k$ starting from $g(d)$. Of course this could be
possible only through a very different design approach.

While our development has demonstrated achievable bounds on the
embedding design, fundamental lower bounds would also be
desirable. For example,~\cite{alon2003problems} demonstrates a lower
bound on the number of measurements required to satisfy the J-L lemma,
which is improved in~\cite{larsen2014johnson} for linear
embeddings. Similarly,~\cite{yi2015binary} has demonstrated some lower
bounds on the rate of binary embeddings. However, neither of these
results account for a distance map and for preserving different ranges
with different accuracy.

Of course, part of our goal is to design representations that
accurately represent signal geometry between pairs of signals, i.e.,
distances and inner products, without requiring both signals to be
present while the representation is computed. These representations
should be easy to compute and easy to compute with, i.e., should
provide straightforward mechanisms to compute the necessary geometric
quantities. While we are using embeddings as our mechanism, it is not
necessarily the only or the optimal approach to the problem.

As a parallel path, consider basis and frame expansions of signals, which
provide straightforward mechanisms to represent signals. Using those representations 
in classical signal processing applications, we can control, for
example, the approximation error in the representation, the
rate-distortion performance of the quantized representation, and the
representation accuracy in certain subspaces, according to the
application requirements. We desire to have the same control on the
signal geometry and not on the signals themselves. In that sense, the
distortion reflects the accuracy of representing distances and
different ranges of distances. Thus, for example, we would like to
control the rate-distortion performance of a quantized distance
representation, given the range of distances we are interested
in. General representation and coding methods, as well as
representation complexity and rate-distorion bounds are still open for
this problem.

Fast computation is also desirable in our methods. While
dimensionality reduction does reduce computation in practice, it does
not improve the asymptotic behavior of methods such as nearest
neighbors. However, there are obvious connections between our work and
LSH~\cite{indyk98stoc,Datar04_LSH,LSH}, which are definitely worth
exploring. LSH could be trivially used as a layer on top of the
embedding, as a separate process to speed up computation. A more
interesting approach would be a combination of the two, since many of
the fundamental concepts and the resulting algorithmic steps are
similar. In some sense, the ideal LSH is also a distance embedding
into a discrete space, in which with very high probability $g(d)=0$ if
$d<r$ and $g(d)\ne 0$ if $d>R$ for some $r<R$.

More generally, we are interested in information representation and
coding when the end goal is a function computation $g(\vx,\vy)$ and when
one of the inputs is not available at the encoder. The information and
the corresponding distortion is measured at the function output. The
embeddings considered in this work represent a special case given by 
$g(\vx,\vy)=g(\|\vx-\vy\|)$. Of interest are more general functions
common in machine learning, such as classifiers and estimators. Some
fundamental bounds and coding schemes have been developed using the
chromatic entropy of the function, e.g.,
see~\cite{orlitsky2001coding,doshi2010functional}. However, these
techniques operate on variables $\vx$ and $\vy$ drawn from a discrete
alphabet. Even with discrete sources, they require the construction of
a large graph that grows with the size of the alphabet.  Such an
approach is prohibitive even in simple problems, such as the distance
representation discussed in this paper.

Distributed functional coding~\cite{misra2011distributed}, is a more
practical and promising alternative for continuous
sources. Unfortunately, distance functions do not satisfy the
conditions for optimality in the proposed solution---specifically, an 
``equivalence-free'' property. This will most definitely also be 
the case for a number of machine learning algorithms. Furthermore, for
such algorithms an explicit differentiable functional form, as
required for the encoding, might not be easily available. For such
functions, an embedding-based coding might be more appropriate. Still,
it remains to be seen whether embeddings that preserve function computation,
other than geometry, are even possible.

%% file: app_proofcontinuous.tex
\section{Proof of Theorem~\ref{thm:emb_continuous}}
\label{app:proof_continuous}
\begin{proof}
  First, we consider two balls of radius $r$ with centers \vx\ and \vy,
  denoted \ball{r}{\vx}\ and \ball{r}{\vy}, respectively. For any vector
  pair $\vx'\in\ball{r}{\vx}$, $\vy'\in\ball{r}{\vy}$ we have
  \begin{align}
    |d(\vx',\vy')-d(\vx,\vy)|&\le 2r\label{eq:tri_ineq_orig}\\
    \Rightarrow |d(f(\vx'),f(\vy'))-d(f(\vx),f(\vy))|&\le 2K_fr\label{eq:tri_ineq_embedding}\\
    \mathrm{and~} |g(d(\vx',\vy'))-g(d(\vx,\vy))|&\le 2K_gr\label{eq:tri_ineq_map}
  \end{align}
  This follows by the triangle inequality and the properties of Lipschitz
  continuity.

  Starting with~\eqref{eq:tri_ineq_embedding} and
  using~\eqref{eq:general_embedding} and \eqref{eq:tri_ineq_map} we can
  derive
  \begin{align}
    d(f(\vx'),f(\vy'))\le& d(f(\vx),f(\vy))+2K_fr\\\le&
    (1+\delta)g(d(\vx,\vy))+2K_fr+\epsilon\\
    \le& (1+\delta)g(d(\vx',\vy'))\nonumber\\&+(1+\delta)2K_gr+2K_fr+\epsilon
  \end{align}
  and
  \begin{align}
    d(f(\vx'),f(\vy'))\ge& d(f(\vx),f(\vy))-2K_fr\\\ge&
    (1-\delta)g(d(\vx,\vy))-2K_fr-\epsilon\\
    \ge&
    (1-\delta)g(d(\vx',\vy'))\nonumber\\&-(1-\delta)2K_gr-2K_fr-\epsilon\\
    \ge&(1-\delta)g(d(\vx',\vy'))\nonumber\\&-(1+\delta)2K_gr-2K_fr-\epsilon
  \end{align}
  i.e.,
  \begin{multline}
    (1-\delta)g(d(\vx',\vy'))-(1+\delta)2K_gr-2K_fr-\epsilon \\\le
    d(f(\vx'),f(\vy')) \le \\(1+\delta)g(d(\vx',\vy'))+(1+\delta)2K_gr+2K_fr+\epsilon  
  \end{multline}
  Setting $r=\frac{\alpha}{(1+\delta)2K_g+2K_f}$ and
  $\widetilde{\epsilon}=\epsilon+\alpha)$ for some $\alpha$, we
  obtain that the final embedding bound
  \begin{align}
    (1-\delta)g(d(\vx',\vy'))-\widetilde{\epsilon} \le
    d(f(\vx'),f(\vy')) \le (1+\delta)g(d(\vx',\vy'))+\widetilde{\epsilon}\label{eq:ineq_balls}
  \end{align}
  holds with probability
  $1-ce^{-Mw(\delta,\widetilde{\epsilon}-\alpha)}$.

  Using the union bound on the $C_{\varepsilon}^{\sS}$ balls that
  cover the signal set with Kolmogorov entropy $E_r^{\sS}$, it follows
  that~\eqref{eq:ineq_balls} holds with probability greater than
  $1-ce^{2E_r^{\sS}-Mw(\delta,\widetilde{\epsilon}-\alpha)}$, which
  decays exponentially with $M$, as long as $M=O(E_r^{\sS})$.
\end{proof}

%% file: app_proofdiscontinuous.tex
\section{Proof of Theorem~\ref{thm:emb_discontinuous}}
\label{app:proof_discontinuous}
\begin{proof}  
  For a single value of $T$, given $M$ measurements, we can use
  Hoeffding's inequality to upper bound the probability that more than
  $P_T(1+c_0)M$ measurements will be exactly $T$-part Lipschitz
  over a single ball \ball{r/2}{\vx}:
  \begin{align}
    P(\mbox{more than $P_T(1+c_0)M$
      measurements are exactly $T$-part Lipschitz})\le e^{-2c_0^2M}.
  \end{align}
  For each $T$, each of those measurements will partition the ball to
  $T$ sets. We set a $T_{\mathrm{max}}$, denoting the level beyond
  which the probability that a function $f_m(\cdot)$ is $T$-part Lipschitz
  is negligible. Thus, using the union bound, a lower bound on the
  probability that for all $T$, at most $P_T(1+c_0)M$ measurements are
  exactly $T$-part Lipschitz continuous is equal to
  $1-T_{\mathrm{max}}e^{-2c_0^2M}-P_F$, where
  $P_F=(\sum_{T=T_{\mathrm{max}}+1}^\infty P_T)$ is considered
  negligible.
  
  Therefore, the embedding partitions the ball into at most
  \begin{align}
    \mathrm{\#~of~Sets}&\le \prod_{T=1}^{T_{\mathrm{max}}}
    T^{P_T(1+c_0)M}=e^{\sum_{T=1}^{T_{\mathrm{max}}}P_T(1+c_0)M\log T}=
    e^{c_1M}
  \end{align}
  sets, with probability greater than
  $1-T_{\mathrm{max}}e^{-2c_0^2M}-P_F$, where
  $c_1=\sum_{T=1}^{T_{\mathrm{max}}}P_T(1+c_0)\log
  T=\sum_{T=2}^{T_{\mathrm{max}}}P_T(1+c_0)\log T$, since $\log 1
  =0$. Note that $P_T$ concentrates to lower $T$'s as $r$ decreases
  and, therefore, we expect $c_1$ to decrease as $r$ decreases.

  Since the ball has radius $r/2$, i.e., diameter $r$, each set of its
  partition also has the same diameter. In other words, if we pick any
  point in each set of the partition and call it the ``center'' of the
  set, all other points of the set are within $r$ of the center. Thus
  we can repeat the argument of the previous section but on the
  $e^{c_1M}$ set centers produced by each of the $C_{r/2}^\sS$ balls
  that constitute the $r/2$-covering of the set; a total of
  $e^{2E_{r/2}^\sS+c_1M}$ centers. Thus, with probability
  $1-(ce^{2E_{r/2}^\sS+c_1M-Mw(\delta,\widetilde{\epsilon}-\alpha)}-T_{\mathrm{max}}e^{-2c_0^2M}-P_F)$,
  the embedding satisfies~\eqref{eq:ineq_balls}:
  \begin{align}
    (1-\delta)g(d(\vx,\vy))-\widetilde{\epsilon} \le
    d(f(\vx),f(\vy)) \le (1+\delta)g(d(\vx,\vy))+\widetilde{\epsilon}
  \end{align}
  for all \vx\ and \vy\ in \sS, where
  $r=\frac{\alpha}{(1+\delta)2K_g+2K_f}$
\end{proof}

Of course, an analysis along this line can be extended to
multidimensional discontinuous functions for which the continuity in
each dimension cannot be considered independently of the other
dimension. In this case, we need to considering the $T$-part
Lipschitz continuity property in multiple dimensions and perform the
same analysis. In the interest of space, we do not describe this
extension.

%% file: app_proofdesign.tex
\section{Proof of Theorem~\ref{thm:point_embedding}}
\label{app:proof_pointembedding}
\begin{proof}
  We consider the single coefficient $y=h(\langle \va,\vx\rangle+w)$,
  the pair of signals \vx\ and $\vx'$ at distance
  $d=d_{\sS}(\vx-\vx')$ apart, and their (signed) projected distance
  $l=\langle \va, \vx-\vx' \rangle$. Studying how the mapping operates
  on this pair provides the basis for how the mapping operates on the
  whole set of signals, in a manner similar
  to~\cite{dasgupta03rsa,achlioptas03css,BarDavDeV::2008::A-Simple-Proof,B_TIT_12}.

  Conditioned on $l$, the squared difference of the signals' mapping
  has expected value over $w$ equal to
  \begin{align}
    E\{(y-y')^2|l\}&=\int_{0}^{1}(h(u+w)-h(u+l+w))^2f_w(w)dw\\
    &=\int_{0}^{1}h^2(u+w)+h(u+l+w)^2-2h(u+w)h(u+l+w)dw\\
    &=2\left(R_{h}(0)-R_{h}(l)\right),
  \end{align}
  with the last equality following from the shift-invariance of the
  autocorrelation.

  Thus, as a function of $d$, the expected value of the squared
  difference is
  \begin{align}
    E\{(y-y')^2\}&=\int_{-\infty}^{+\infty}E\{(y-y')^2|l\}f_l(l|d)dl\\
    &=\int_{-\infty}^{+\infty}2\left(R_{h}(0)-R_{h}(l)\right)f_l(l|d)dl\\
    &=2-2\int_{-\infty}^{+\infty}R_{h}(l)f_l(l|d)dl
  \end{align}
  Using Parseval's theorem and the characteristic function of
  $f_l(l|d)$, denoted using $\phi_l(\xi|d)$ we obtain
  \begin{align}
    E\{(y-y')^2\}&=2\left(\sum_k(|H_k|^2-|H_k|^2\phi_l(2\pi k|d))\right)\\
    &=2\sum_k|H_k|^2(1-\phi_l(k|d))\label{eq:final_exp}=g(d),
  \end{align}
  where $g(d)$ is the distance map~\eqref{eq:dist_map}.

  We assume that $h(t)$ is bounded and
  denote its range using $\bar{h}=\sup_th(t)-\inf_th(t)$. Thus, the
  squared difference of any measurement is bounded, i.e.,
  $(y-y')^2\in[0,\bar{h}^2]$. Using Hoeffdings inequality, it follows
  that for $M$ measurements
  \begin{align}
    P\left(\left|\frac{1}{M}\sum_m\left(y_m-y_m\right)^2-g(d)\right|\ge\epsilon\right)=P\left(\left|\frac{1}{M}\left\|\vy-\vy'\right\|_2^2-g(d)\right|\ge\epsilon\right)\le2e^{-2M\frac{\epsilon^2}{\bar{h}^4}}.
  \end{align}
  As we describe in Section~\ref{sec:embedd-conc-meas}, using the
  union bound on a set \sS\ of $Q$ points, i.e., at most $Q^2/2$ point
  pairs, the embedding guarantee in the theorem follows.
\end{proof}

%% file: app_proofsubadditive.tex
\section{Proof of Proposition~\ref{prop:subadditivity}}
\label{sec:proof-subadditivity}
\begin{proof}
  By definition, the distance functions on both spaces satisfy the
  triangle inequality:
  \begin{align}
    d(\vx,\vz)&\le d(\vx,\vy)+d(\vy,\vz)\\
    d(f(\vx),f(\vz))&\le d(f(\vx),f(\vy))+d(f(\vy),f(\vz))\label{eq:tri_embedding}
  \end{align}
  To show that $g(d)$ is subadditive, we pick $\vy$ in the line
  between $\vx$ and $\vz$, such that $d(\vx,\vz)=d$,
  $d(\vx,\vy)=\lambda d$, and $d(\vy,\vz)=(1-\lambda)d$. This can be
  trivially done if \sS\ is a convex
  set. From~\eqref{eq:tri_embedding} and~\eqref{eq:general_embedding} it
  follows that
  \begin{align}
    (1-\epsilon)g(d)-\delta &\le (1+\epsilon)g(\lambda
    d)+\delta+(1+\epsilon)g((1-\lambda) d)+\delta\\
    \Rightarrow  \frac{1-\epsilon}{1+\epsilon}g(d)-3\delta&\le g(\lambda
    d)+g((1-\lambda)d)\\
    \Rightarrow  (1-2\epsilon)g(d)-3\delta&\le g(\lambda
    d)+g((1-\lambda)d)
  \end{align}
  Selecting $a,b$ and $d$ such that $a+b=d$ and setting
  $\lambda=a/(a+b)$ proves that $g(d)$ is
  $(2\epsilon,3\delta)$-subadditive.
\end{proof}

%% file: app_proofbinarycollisionbounds.tex
\section{Probability of Crossing a Universal Quantization Threshold}
\label{sec:prob-cross-univ}
We develop a bound for $P_2$ assuming the measurement matrix \mA\ has
entries drawn from an i.i.d. $\mathcal{N}(0,\sigma^2)$
distribution. In that case, the $m^{th}$ projection of a ball of
radius $r/2$ will have diameter at most $\|\va\|_2r$ where $\|\va\|_2$
is the norm of an $N$-dimensional Gaussian vector with variance
$\sigma^2$. In other words, the worst-case projected diameter is
proportional to a $\chi$-distributed random variable with $N$ degrees
of freedom. We denote its distribution using 
\begin{align}
  f_{l|r}(l)=\frac{1}{\sigma r}\chi_N\left(\frac{l}{\sigma r}\right),
\end{align}
where $\chi_N(\cdot)$ denotes the density of the $\chi$ distribution
with $N$ degrees of freedom.

The probability that the projection of the ball will cross a
quantization boundary and, therefore, will become exactly 2-part Lipschitz continuous,
i.e., $P_2$ can then be upper bounded using
\begin{align}
  P_2&\le \int_0^\Delta \frac{l}{\Delta}f_{l|r}(l)dl +
  \int_\Delta^{\infty}f_{l|r}(l)dl=\frac{\sigma
    r}{\Delta}\int_0^{\frac{\Delta}{\sigma r}} l \chi_N(l) dl
  +\int_{\frac{\Delta}{\sigma r}}^{\infty} \chi_N(l) dl\\ &\le
  \frac{\sigma r}{\Delta}\int_0^{\infty} l \chi_N(l) dl
  +\int_{\frac{\Delta}{\sigma r}}^{\infty} \chi_N(l) dl=\frac{\sigma
    r}{\Delta} E\{\chi_N\} +\int_{\frac{\Delta}{\sigma r}}^{\infty}
  \chi_N(l) dl\\ &\le \frac{\sigma
    r}{\Delta}\frac{\sqrt{2}\Gamma((N+1)/2)}{\Gamma(N/2)}+
  \gamma\left(\frac{N}{2},\left(\frac{\Delta}{2\sigma
    r}\right)^2\right),
  \label{eq:bound_on_P2}
\end{align}
where $\Gamma(\cdot)$ is the Gamma function, $\gamma(s,x)$ is the
regularized upper incomplete gamma function, and
$\gamma\left(\frac{N}{2},\left(\frac{x}{2}\right)^2\right)$ is the
tail integral of the $\chi$ distribution with $N$ degrees of freedom
(i.e., the distribution of the norm of a $N$-dimensional standard
Gaussian vector). In other words,
\begin{align}
   \gamma\left(\frac{N}{2},\left(\frac{\Delta}{2\sigma
     r}\right)^2\right) = P(\|a\|_2r\ge\Delta)
\end{align}

Using a simple concentration of measure argument we can obtain
\begin{align}
  P(\|\va\|_2\ge (1+\beta)\sigma\sqrt{N})&\le e^{-\beta^2
    N/6}\\ \Rightarrow P(\|\va\|_2r\ge (1+\beta)\sigma\sqrt{N}r)&\le
  e^{-\beta^2 N/6}.
\end{align}
Substituting $\beta=\frac{\Delta}{\sigma r \sqrt{N}}-1$, it follows
that 
\begin{align}
  P(\|a\|_2r\ge\Delta)\le e^{-\left(\frac{\Delta}{\sigma r
      \sqrt{N}}-1\right)^2N/6}
 \label{eq:norm_tail_bound} 
\end{align}
Using the Gautschi-Kershaw inequality on the ratio of Gamma functions
in~\eqref{eq:bound_on_P2} and the tail bound
in~\eqref{eq:norm_tail_bound}, $P_2$ is bounded by
\begin{align}
  P_2\le \frac{\sigma r \sqrt{N+1}}{\Delta} +
    e^{-\left(\frac{\Delta}{\sigma r \sqrt{N}}-1\right)^2N/6} :=\overline{P}_2,
\end{align}
which becomes meaningful when $r<\frac{\Delta}{\sigma\sqrt{N+1}}$, and
approaches 0 as $r$ decreases.

Thus, $c_1\le \overline{P}_2(1+c_0)\log 2$ and the probability that the embedding
does not hold is upper bounded by
\begin{align}
  ce^{2E_{r/2}^\sS+c_1M-Mw(\delta,\epsilon)}+
  T_{\mathrm{max}}e^{-2c_0^2M}+P_F\le
  e^{2E_{r/2}^\sS+M\overline{P}_2(1+c_0)\log 2-2M\epsilon^2}+
  2e^{-2c_0^2M}
\end{align}
which decreases exponentially with $M$, as long as
$2\epsilon^2>\overline{P}_2(1+c_0)\log 2$, allowing the embedding
error $\epsilon$ to approach 0 with appropriate choice of $r$. Note
that as $r$ decreases, $P_2$ decreases approximately linearly whereas
$E_{r/2}^\sS$ increases as $\dim^\sS\cdot\log(1/r)$, where $\dim^\sS$ is the Kolmogorov
dimension of the set $\sS$:
\begin{align}
  \dim^\sS=\lim_{r\rightarrow 0}\frac{\log
    C_r^\sS}{\log(1/r)}=\lim_{r\rightarrow 0}\frac{\log
    E_r^\sS}{\log(1/r)}.
\end{align}